\documentclass[pre,twocolumn,amsmath,amsfonts,floatfix,superscriptaddress]{revtex4-1}
\usepackage{graphicx,amssymb}
\usepackage{color}
\usepackage{bm} 
\newcommand{\rev}[1]{\textcolor{black}{{#1}}}
\newcommand{\rrev}[1]{\textcolor{black}{{#1}}}
\begin{document}

\title{Stable glassy configurations of the Kob-Andersen model using swap Monte Carlo}

\author{Anshul D. S. Parmar}

\affiliation{Laboratoire Charles Coulomb (L2C), Universit\'e de Montpellier, CNRS, 34095 Montpellier, France.}

\author{Benjamin Guiselin}

\affiliation{Laboratoire Charles Coulomb (L2C), Universit\'e de Montpellier, CNRS, 34095 Montpellier, France.}

\author{Ludovic Berthier}

\affiliation{Laboratoire Charles Coulomb (L2C), Universit\'e de Montpellier, CNRS, 34095 Montpellier, France.}

\affiliation{Department of Chemistry, University of Cambridge, Lensfield Road, Cambridge CB2 1EW, United Kingdom.}

\email{ludovic.berthier@umontpellier.fr}

\date{\today}

\begin{abstract}
The swap Monte Carlo algorithm allows the preparation of highly stable glassy configurations for a number of glass-formers, but is inefficient for some models, such as the much studied binary Kob-Andersen (KA) mixture. We have recently developed generalisations to the KA model where swap can be very effective. Here, we show that these models can in turn be used to considerably enhance the stability of glassy configurations in the original KA model at no computational cost. We \rev{successfully} develop several numerical strategies both in and out of equilibrium to achieve this goal and show how to optimise them. 
\rev{We provide several physical measurements indicating that the proposed algorithms considerably enhance mechanical and thermodynamic stability in the KA model, including a transition towards brittle yielding behaviour.  Our results thus pave the way for future studies of stable glasses using the KA model.}

\end{abstract}

\maketitle

\section{Introduction}

\label{sec:introduction}

Computer simulations of supercooled liquids and glasses play an important role to link their physical properties to the structure and dynamics at the microscopic scale~\cite{berthier2011theoretical}. Structural relaxation and equilibration are however so slow that simulating glass-forming liquids is generally difficult. This problem was recently solved for a broad \rev{(but incomplete)} class of model glass-formers using the swap Monte Carlo algorithm~\cite{berthier2016equilibrium,ninarello2017models,berthier2019efficient}. For some models, an equilibration speedup larger than $10^{11}$ was achieved, opening the door to direct comparisons between numerical and experimental work~\cite{fullerton2017density,berthier2017configurational,berthier2019zero,scalliet2019nature,wang2019low,ozawa2019does,berthier2019finite,khomenko2020depletion,guiselin2020random}.  

The Kob-Andersen (KA) model is a binary mixture of Lennard-Jones particles devised to describe the generic physical properties of simple metallic glasses~\cite{kob1995testing}. For this well-studied model, the swap Monte Carlo algorithm is inefficient as the swap of unlike species is almost always rejected~\cite{flenner2006hybrid}. Therefore, the simulation of low-temperature properties of the KA model requires alternative methods, such as parallel tempering~\cite{coslovich2018dynamic}, simulations on graphic cards~\cite{coslovich2018dynamic,schroder2020solid}, ghost particle insertion~\cite{turci2017nonequilibrium}, Wang-Landau algorithm~\cite{faller2003density}, transition path sampling~\cite{jack2011preparation,turci2017nonequilibrium}, physical vapor deposition~\cite{lyubimov2013model}, oscillatory shear~\cite{das2018annealing,priezjev2018molecular,bhowmik2020thermodynamic}.  \rev{However, none of these attempts could provide the type of speedup that the swap Monte Carlo has provided for the models mentioned before. There is thus a clear need to further develop computational algorithms to produce more stable glassy configurations of the KA model.} 
  
Recently, we introduced generalised versions of the KA model (called KA$_1$ and KA$_2$ models) which are very similar to the original KA model, and for which the swap Monte Carlo algorithm is very efficient~\cite{parmar2020ultrastable}. The strategy relies on introducing a small amount of additional species to the binary KA mixture to enhance the swap efficiency. This strategy will allow the investigation of properties of simple metallic glasses down to the experimental glass transition, but not for the KA model itself. In this work, we demonstrate that the production of very stable configurations within the KA$_1$ model can in turn be used to produce stable glassy configurations of the original KA model as well, for a modest computational effort. \rev{Because many groups have developed numerical expertise and analysis tools for the KA model, our results pave the way for future studies of stable glass physics within the KA model.}

The manuscript is organised as follows. In Sec.~\ref{sec:models} we define the various glass models we used. In Sec.~\ref{sec:reweighting}, we use histogram reweighting techniques to measure equilibrium properties of the KA model. In Sec.~\ref{sec:annealing} we present two annealing procedures to prepare stable configurations of the KA model. In Sec.~\ref{sec:stability} we quantify the stability of the obtained configurations. We conclude in Sec.~\ref{sec:conclusion}.

\section{Models} 

\label{sec:models}

We consider mixtures of particles $i=1,..,N$ of different species characterized by a number $\omega_i\in[0;1]$. The interaction potential between two particles $i$ and $j$ is
\begin{equation}
v(r_{ij};\omega_i,\omega_j) = 4\epsilon_{\omega_i\omega_j} \left[ \left( \frac{\sigma_{\omega_i\omega_j} }{r_{ij}}  \right)^{12} - \left( \frac{ \sigma_{\omega_i \omega_j} }{r_{ij}} \right)^{6} \right],
\label{eqn_potential}
\end{equation}
which depends on the distance $r_{ij}$ between the two particles, on the interaction strength $\epsilon_{\omega_i\omega_j}$, and on the cross-diameter $\sigma_{\omega_i\omega_j}$. The potential is truncated and shifted at the cutoff distance $r_{{\rm cut},\omega_i\omega_j} = 2.5 \sigma_{\omega_i\omega_j}$. 

We focus on two related models. The first one is the standard Kob-Andersen model~\cite{kob1995testing} which is a 80:20 binary mixture of $N_A$ particles of type A (with $\omega_i=1$) and $N_B$ particles of type B (with $\omega_i=0$). The interaction parameters are: $\epsilon_{AB}/\epsilon_{AA}=1.5$, $\epsilon_{BB}/\epsilon_{AA}=0.5$, $\sigma_{AB}/\sigma_{AA}=0.8$ and $\sigma_{BB}/\sigma_{AA}=0.88$. Energies and lengths are expressed in units of $\epsilon_{AA}$ and $\sigma_{AA}$ respectively\rev{, and the Boltzmann constant is set to unity}. We denote $\mathcal{H}_{\rm KA}[\bm{r}^N]$ the corresponding Hamiltonian \rev{of the KA model.}

We also consider an extended version of the KA model (KA$_1$) by introducing a small fraction $\delta=N_C/(N_A+N_B)$ of particles of type $C$ interpolating continuously between $A$ and $B$ particles. More precisely, $C$ particles are characterized by a uniform distribution of $\omega_i \in ]0;1[$, \rev{while $A$ (resp. $B$) particles are still associated to the type $\omega_i=1$ (resp. $\omega_i=0$)}.
The Hamiltonian of the KA$_1$ model is 
\begin{equation}
\mathcal{H}_1[\bm{r}^N]=\sum_{i<j} v(r_{ij};\omega_i,\omega_j),
\label{eqn_E1}
\end{equation}
with the additional interaction parameters
\begin{equation}
\begin{aligned}
X_{1\omega_i}&=\omega_i X_{AA}+(1-\omega_i)X_{AB},\\
X_{0\omega_i}&=\omega_i X_{AB}+(1-\omega_i)X_{BB},\\
X_{\omega_i\omega_j}&=\omega_{ij} X_{AA}+(1-\omega_{ij})X_{BB},
\end{aligned}
\label{eqn_parameters}
\end{equation}
where $X=\sigma,\epsilon$ and $\omega_{ij}=(\omega_i+\omega_j)/2$~\cite{parmar2020ultrastable}. We also define the Hamiltonian $\mathcal{H}_0[\bm{r}^N]$ the system would have if $C$ particles with $\omega_i \leq 0.2$ (resp. $\omega_i>0.2$) were taken as B (resp. A) particles. Thus, $\mathcal{H}_0$ is the Hamiltonian of the corresponding KA model, given by Eq.~(\ref{eqn_E1}) with $\omega_i$ replaced by $\omega_i^\prime=1-\theta(1-\omega_i/0.2)$, with $\theta(x)$ the Heaviside function. Finally, we define 
\begin{equation}
\mathcal{W}=\mathcal{H}_1-\mathcal{H}_0
\end{equation} 
as the energy difference between the KA$_1$ and the KA energies for a given configuration of the KA$_1$ model. 

We study the KA$_1$ model with $N_C=5$, $N_A=800$ and $N_B=200$ (so that $\delta = 0.5\%$), at number density $\rho=1.2$ \rev{under periodic boundary conditions}. We use 9 times larger systems for the rheology, see below. The model is studied using the swap Monte Carlo algorithm. \rev{With probability $p=0.2$, the identity of particle $i$ is exchanged with the one of particle $j$, both particles being randomly chosen. Otherwise, with probability $1-p$, a standard translational move is performed in which the position $\bm{r}_i$ of particle $i$ is incremented by a random displacement $\delta\bm r_i$ drawn in a cube of linear size $0.15$ centered around the origin~\cite{berthier2007monte}. Both moves are accepted according to the Metropolis rule. We fix the position of the center of mass.} 

Due to the large difference in diameters between A and B particles, swap moves are inefficient in the KA model~\cite{flenner2006hybrid}, whereas the introduction of a small fraction of $C$ particles makes swap moves possible and results in a much faster relaxation~\cite{parmar2020ultrastable}. The structural relaxation time $\tau_{\alpha}$ of the system is defined as the time value at which the self-part of the intermediate scattering function calculated for the whole system, with a wave number corresponding to the first peak of the total structure factor, decays to the value $1/e$. It is expressed in units of Monte Carlo (MC) steps, where 1 MC step corresponds to $N$ attempted moves. The lowest temperature for which we can ensure equilibration in the KA model is $T \simeq 0.415$, whereas for the KA$_1$ we can reach $T \simeq 0.36$ for a comparable numerical effort of $10^{8}$ Monte Carlo steps. In terms of $\tau_\alpha$, this represents a speedup factor of more than $10^2$ over the standard KA model at the lowest temperature.

\begin{figure*}
\includegraphics[scale=0.38]{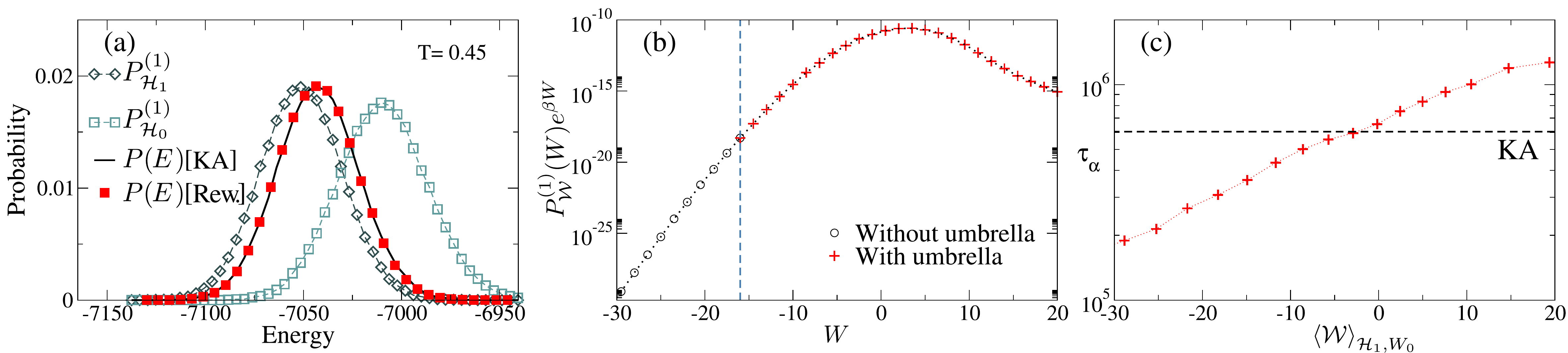}
\caption{(a) Probability distributions $P_{\mathcal{H}_1}^{(1)}$, $P_{\mathcal{H}_0}^{(1)}$ of energies $\mathcal{H}_1$ and $\mathcal{H}_0$ in the KA$_1$ system at $T=0.45$ along with the reweighted probability distribution $\rev{P}$ [Rew.] of the energy for the KA model obtained from Eq.~(\ref{eqn_proba_rew}). The probability distribution $\rev{P}$ has also been directly measured in the KA model [KA] to check the quality of the reweighting procedure. (b) Plot of $P_\mathcal{W}^{(1)}(W)e^{\beta W}$ with (red) and without (black) umbrella sampling. The dashed line marks the limit of $W$ in unbiased Monte Carlo simulations of the KA$_1$ model. (c) Relaxation time $\tau_\alpha$ of the system in the different umbrella simulations as a function of $\langle \mathcal{W}\rangle_{\mathcal{H}_{1,W_0}}$, the average value of $\mathcal{W}$.}
\label{fig1}
\end{figure*}

\section{Reweighting equilibrium distributions}

\label{sec:reweighting}

In this section, we show how to compute the thermodynamic properties of the KA model from simulations of the KA$_1$ model using reweighting methods~\cite{newman1999monte}. Since the swap Monte Carlo algorithm efficiently thermalises the KA$_1$ model, the use of histogram reweighting can potentially produce thermodynamic properties for the KA model at the low temperatures where only the KA$_1$ model can reach equilibrium.

In particular, we focus on the probability distribution of the energy in the standard KA model
\begin{equation}
\begin{aligned}
P(E)&=\langle\delta(E-\mathcal{H}_{\rm KA})\rangle_{\mathcal{H}_{\rm KA}}\\
&=\frac{\displaystyle \int \mathrm{d}\bm{r}^N \delta(E-\mathcal{H}_{\rm KA} [\bm{r}^N]) e^{-\beta\mathcal{H}_{\rm KA}[\bm{r}^N]}}{\displaystyle \int \mathrm{d}\bm{r}^N e^{-\beta\mathcal{H}_{\rm KA}[\bm{r}^N]}},
\end{aligned}
\label{eqn_proba_0}
\end{equation}
with $\delta(x)$ the delta function and $\langle \cdots \rangle_{\mathcal{H}_{\rm KA}}$ the thermodynamic average at inverse temperature $\beta=T^{-1}$ for the Hamiltonian $\mathcal{H}_\mathrm{KA}$.

It is useful to rewrite Eq.~(\ref{eqn_proba_0}) using quantities defined within the KA$_{1}$ model,
\begin{equation}
\begin{aligned}
P(E)&=\frac{\displaystyle \int \mathrm{d}\bm{r}^N \delta(E-\mathcal{H}_0\rev{[\bm{r}^N]}) e^{-\beta\mathcal{H}_1\rev{[\bm{r}^N]}+\beta \mathcal{W}\rev{[\bm{r}^N]}}}{\displaystyle \int \mathrm{d}\bm{r}^N e^{-\beta\mathcal{H}_1\rev{[\bm{r}^N]}+\beta \mathcal{W}\rev{[\bm{r}^N]}}}\\
&=\frac{\langle\displaystyle \delta(E-\mathcal{H}_0) e^{\beta \mathcal{W}}\rangle_{\mathcal{H}_1}}{\langle\displaystyle e^{\beta \mathcal{W}}\rangle_{\mathcal{H}_1}},
\end{aligned}
\label{eqn_proba_1}
\end{equation}
where now $\langle \cdots \rangle_{\mathcal{H}_1}$ stands for the thermodynamic average for the Hamiltonian $\mathcal{H}_1$. We used Eq.~(\ref{eqn_E1}) and the fact that, by definition, $\mathcal{H}_0=\mathcal{H}_\mathrm{KA}$.

Introducing $P_\mathcal{W}^{(1)}(W)=\langle \delta(W-\mathcal{W})\rangle_{\mathcal{H}_1}$ the probability distribution of $\mathcal{W}$ measured in the KA$_1$ model, using the trivial identity  $e^{\beta \mathcal{W}}=\int\mathrm{d}W e^{\beta W}\delta(W-\mathcal{W})$ and the Kolmogorov definition of a conditional probability, Eq.~(\ref{eqn_proba_1}) can be written as
\begin{equation}
P(E) = \frac{\displaystyle \int \mathrm{d}W P_{\mathcal{H}_0|\mathcal{W}}^{(1)}(E|W) P_{\mathcal{W}}^{(1)}(W) e^{\beta W}}{\displaystyle \int \mathrm{d}W P_\mathcal{W}^{(1)}(W) e^{\beta W}}, 
\label{eqn_proba_rew}
\end{equation}
where $P_{\mathcal{H}_0|\rev{\mathcal{W}}}^{(1)}$ is the conditional probability of $\mathcal{H}_0$ given $\mathcal{W}$, measured in the KA$_1$ model. The two distributions in the right-hand side of Eq.~(\ref{eqn_proba_rew}) can be measured in the course of a simulation of the KA$_1$ model and subsequently reweighted to obtain the probability distribution of the energy in the KA model in the left-hand side. Thus, in principle, the properties of the KA model can be obtained without ever performing a simulation of the KA model itself but only working with the KA$_1$ model where swap Monte Carlo works well. 

In Fig.~\ref{fig1}(a), we show the distributions of $\mathcal{H}_{1}$ and 
$\mathcal{H}_{0}$ measured in the KA$_1$ model, for a temperature $T=0.45$ for which the relaxation time of the KA model is $\tau_\alpha/\tau_0 \simeq 2 \times 10^2$, with $\tau_0\simeq 3 \times 10^3$ the relaxation time at the onset temperature \rev{$T_0\simeq 0.70$ of glassy behaviour (corresponding to the appearance of a two-step decay in correlation functions)}. We also show $P(E)$ directly measured in the KA model to assess the validity of the reweighting procedure. The product $P_\mathcal{W}^{(1)}(W) e^{\beta W}$ plays a crucial role in the reweighting scheme as emphasized by Eq.~(\ref{eqn_proba_rew}). However, as shown in Fig.~\ref{fig1}(b), this quantity increases (exponentially) without bounds in the range of $\mathcal{W}$ that is being explored in a direct simulation of the KA$_1$ model. This finding indicates that a direct application of Eq.~(\ref{eqn_proba_rew}) is not possible with this set of data, as the tails of the distributions involved in the various integrands are not appropriately sampled. This limitation becomes increasingly difficult to tackle when $\delta$ increases, which explains why we chose the smallest value $\delta = 0.5\%$ studied in Ref.~\cite{parmar2020ultrastable}. 

To overcome this sampling issue, we need to force the system to visit non-typical, larger values of $\mathcal{W}$. To this end, we use umbrella sampling techniques~\cite{frenkel2001understanding,torrie1977monte}. We perform several simulations of the KA$_1$ model in parallel, each simulation being run with a biased Hamiltonian of the form 
\begin{equation}
\mathcal{H}_{1,W_0}=\mathcal{H}_1+\kappa(\mathcal{W}-W_0)^{2},
\label{eq:umbrella}
\end{equation} 
with $\kappa=0.05$ the strength of the bias, in order to be able to sample values of $\mathcal{W} \simeq W_0 \in [-17.5;20]$. By combining the different umbrella simulations, we can extend the range over which $P_{\mathcal{H}_0|\mathcal{W}}^{(1)}$ and $P_{\mathcal{W}}^{(1)}$ are measured. The latter is obtained by histogram reweighting, as \begin{equation}
P_{\mathcal{W}}^{(1)}(W)=\mathcal{Z}(W_0)P_{\mathcal{W}}^{(1,W_0)}(W)e^{\beta \kappa(W-W_0)^{2}},
\end{equation}
with $P_{\mathcal{W}}^{(1,W_0)}$ the probability density of $\mathcal{W}$ with the bias and $\mathcal{Z}(W_0)$ an unknown normalisation constant~\cite{ferrenberg1988new}. For two consecutive values of $W_0$, the ratio of these normalisation constants can be estimated from the range of overlapping values of the biased probabilities~\cite{hartmann2011large}. Thus, the estimates of $P_{\mathcal{W}}^{(1)}$ from different umbrella simulations can be glued together, and for each bin the most accurate value is kept. In Fig.~\ref{fig1}(b), we show that $P_{\mathcal{W}}^{(1)}(W)\exp(\beta W)$ is now bounded with a maximum for $W \simeq 0$. This implies that the integrals in Eq.~(\ref{eqn_proba_rew}) are dominated by configurations having $\mathcal{W}\simeq 0$, namely KA-like configurations. After umbrella sampling, Eq.~(\ref{eqn_proba_rew}) can now be numerically evaluated to obtain an accurate estimate of $P(E)$, see Fig.~\ref{fig1}(a).

We have shown that thermodynamic properties of the KA model can be obtained from simulations of the KA$_1$ model, which can involve the efficient swap moves. However, these measurements rely on umbrella sampling simulations, and care must be taken that these biased simulations are all performed in equilibrium conditions. To ensure a proper sampling in the umbrella simulations, we measure the relaxation time $\tau_\alpha$ as a function of $\langle \mathcal{W}\rangle_{\mathcal{H}_{1,W_0}}$, the average value of $\mathcal{W}$ [see Fig.~\ref{fig1}(c)]. It turns out that $\tau_\alpha$ increases from its value in the KA$_1$ model to its value in the KA model when $\langle \mathcal{W}\rangle_{\mathcal{H}_{1,W_0}} \simeq 0$, and increases further for positive values. The physical interpretation is that 
the biased KA$_1$ system visits KA-like configurations when $\langle \mathcal{W}\rangle_{\mathcal{H}_{1,W_0}} \simeq 0$, for which the swap algorithm is inefficient, despite the fact that the acceptance rate of swap moves is actually very high. This means that the frequently-accepted swap moves in the biased KA$_1$ model do not accelerate the equilibration.  

Therefore, the strategy devised here does work correctly, and numerical results for the KA model can be obtained without ever simulating it. However it can only be implemented at sufficiently high temperatures, as one needs  to achieve equilibration times close to the one of the KA model itself to implement the reweighting procedure. In other words, at equilibrium, we can measure $P(E)$ using the KA$_1$  model only in a range of temperatures for which it can directly be measured in the KA model as well, as a continuous chain of equilibrium simulations interpolating between KA$_1$ and KA models is needed. There is thus no computational advantage. 

\section{Two annealing procedures}

\label{sec:annealing}

To produce useful results for the KA model using only the KA$_1$ model, one needs to smoothly transform KA$_1$ data into KA ones. If done in fully equilibrium conditions, a bottleneck is necessarily encountered as the final steps involve being in equilibrium within a system close to the KA model. This is always problematic, as the swap Monte Carlo algorithm does not work well in this regime. 

In this section, we again transform KA$_1$ results (which benefit from the swap algorithm) into KA ones (which do not), but relax the constraint that the final configurations are at equilibrium. To this end, we develop two annealing procedures to smoothly transform in a finite amount of time very stable KA$_1$ configurations into KA ones. The hope is that the gain in stability in the first steps is not completely lost during the annealing procedure. 

In method I, we perform simulations with the Hamiltonian $\mathcal{H}_{1,W_0}$ and we linearly increase the value of the bias $W_0$ up to $W_0 = 0$ (the system is then close to the KA model) in a total number of Monte Carlo steps $t_\mathrm{MC}$. \rev{Initially, the value of $W_0$ is set to the instantaneous value of $\mathcal{W}$ in the initial configuration.} We then switch the Hamiltonian to ${\cal H}_0$, which is equivalent to treating the final configuration as a bona fide KA configuration.  In this method, the KA$_1$ model is then gradually biased using the umbrella sampling Hamiltonian in Eq.~(\ref{eq:umbrella}) towards the KA model.

In method II, we \rev{do not rely on umbrella sampling and we} always use the \rev{bulk} KA$_1$ Hamiltonian\rev{. We} gradually convert the minority species C particles into A or B particles, thus achieving the desired ${\cal H}_1 \to {\cal H}_0$ annealing. In practice, we run simulations with Hamiltonian $\mathcal{H}_{1}$ and at each MC step, with probability $p_\omega=1/50$, we pick up at random one C particle and we increase (resp. decrease) its variable $\omega_i$ by a small increment $d\omega$ if initially $\omega_i>0.2$ (resp. $\omega_i\leq 0.2$). \rev{Otherwise, with probability $1-p_\omega$, we perform translational or swap moves according to the procedure presented in Sec.~\ref{sec:models}.} The increment $d\omega$ is chosen so that after an average number of $t_\mathrm{MC}$ MC steps, the C particles are all converted into either A or B particles. We can then switch the Hamiltonian to ${\cal H}_0$, which is again equivalent to treating the final configuration as a bona fide KA configuration. \rev{We should stress that, even though the annealing methods may look artificial regarding other preparation protocols (like gradual cooling or aging), the only thing that matters is that a genuine amorphous KA glass is eventually obtained through the algorithm.} 

In both methods, we transform KA$_1$ into KA configurations at constant temperature. More complicated annealing schemes could involve changing other parameters as well~\cite{van1987simulated}, but we leave them for future work. To test our methods, and compare the relative efficiency of all schemes, we decided to use similar computational effort (i.e. similar CPU times) for all configurations, with a maximum walltime of 2 weeks (corresponding to $10^8$ Monte Carlo steps).

First, we prepared a series of equilibrium and glassy configurations of both KA and KA$_1$ models. For the KA model, equilibration is ensured down to $T=0.415$, and for the KA$_1$ model down to $T=0.36$. To produce glassy configurations at even lower temperatures, we quenched several configurations at these final temperatures to a range of lower temperatures, down to $T=0.30$. The aging time for each of these glasses is $t_w = 10^8$. 

The equilibrium (for $T \geq 0.36$) and aged (for $T < 0.36$) KA$_1$ configurations are then slowly annealed using methods I and II towards KA configurations. We used annealing times $t_{MC}$ from $9 \times 10^5$ to $7.5 \times 10^7$, to keep the longest simulations to at most $10^8$ Monte Carlo steps and to ensure a fair comparison between all protocols. To improve the statistics, we performed 30 independent simulations for each temperature. 

As a result of the annealing methods, we obtained an ensemble of KA configurations at various temperatures, whose stability we can compare to direct simulations of the KA model over a similar preparation timescale. For the KA model, we used either equilibrium configurations for $T\geq 0.415$ or configurations aged for $10^8$ Monte Carlo steps for $T<0.415$.   

\section{Stable glassy KA configurations}

\label{sec:stability}

In this section we analyse the KA configurations produced by the annealing methods I and II, and by direct aging with the KA Hamiltonian using various physical quantities. 

\subsection{Inherent structure energies}

\begin{figure}
\includegraphics[width=8.5cm]{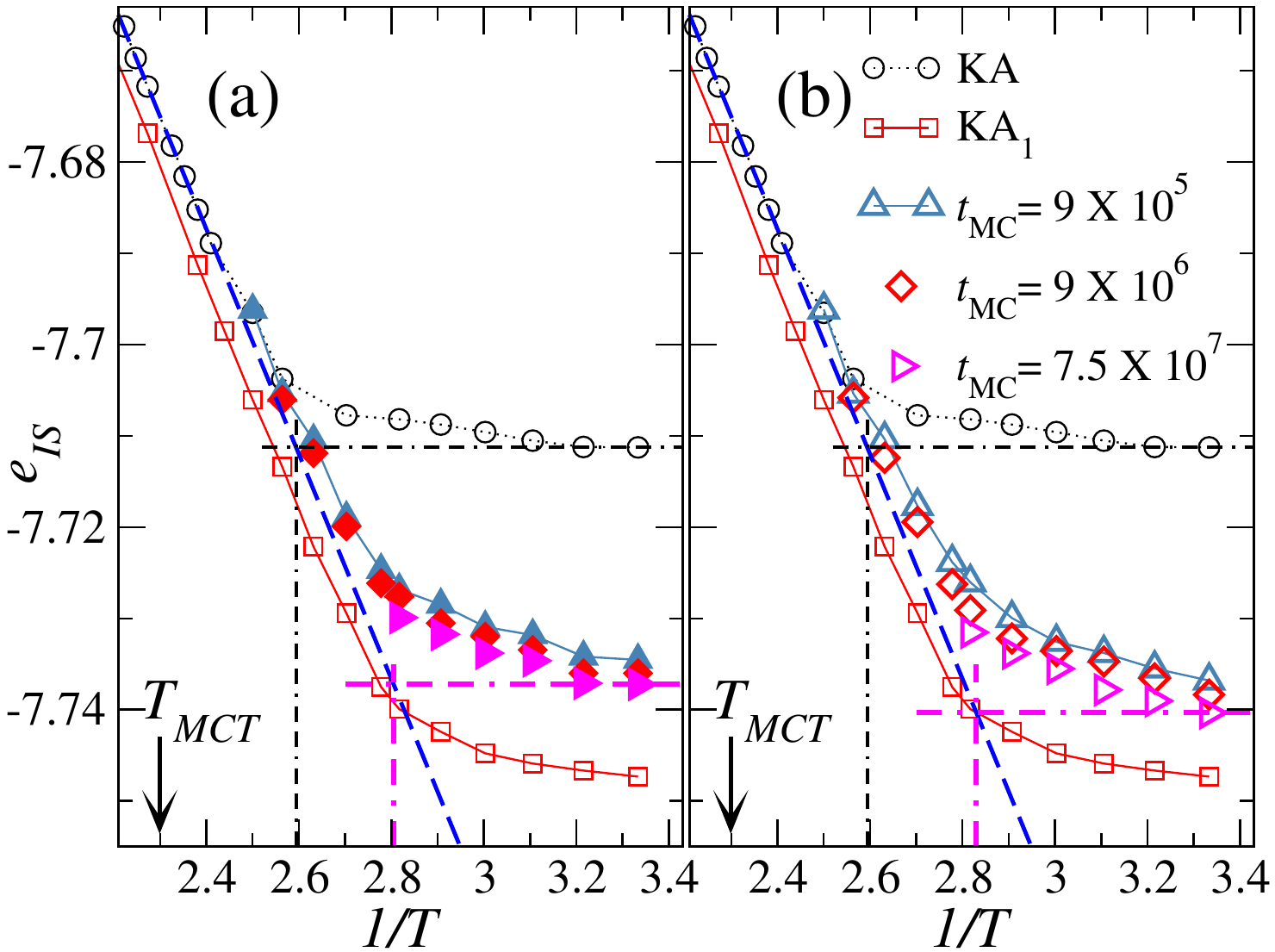}
\caption{Average inherent structure energy per particle $e_\mathrm{IS}$ for the KA and KA$_{1}$ models as a function of inverse temperature. The dashed blue line corresponds to \rev{$e_{IS}^{\rm eq} = a/T+b$} which fits the equilibrium data for the KA model. The IS energies are obtained thanks to method I (panel a) and II (panel b) at different rates. Fictive temperatures are determined via the dashed-dotted lines.}
\label{fig2}
\end{figure}

Our first strategy to quantify the stability of the KA configurations is to quench them to $T=0$ and to record the inherent structure (IS) energy per particle.  In Fig.~\ref{fig2}, we show the average energy of the IS per particle $e_\mathrm{IS}$ for (a) method I and (b) method II as a function of the inverse temperature for three different annealing rates, corresponding to $t_\mathrm{MC}=9\times10^{5}$, $9\times10^6$ or $7.5\times10^{7}$ (and $d\omega=10^{-4}$, $10^{-5}$ or $1.2\times10^{-6}$ for method II). For a given rate, we have checked the influence of the number of C particles and we found that for concentrations larger than $\delta=0.5\%$, higher energy states were reached. This is why we only show results for $\delta= 0.5\%$. In addition, we clearly see that the lower annealing rates give lower IS energies at fixed temperature. 

As a matter of comparison, we show two additional data sets in Fig.~\ref{fig2}. The first one represents the average IS energy of the KA$_1$ model for the set of initial conditions described before. The second one corresponds to the IS energies obtained directly in the KA model, as explained in the previous section. The annealing data clearly lie above the data for the KA$_1$ model, suggesting that during the annealing some of the initial stability gained via the swap Monte Carlo algorithm is lost. However, the annealed states lie much below the IS energies obtained by direct aging in the KA model using a comparable numerical effort. \rrev{The KA configurations obtained by aging a time $t_w=10^8$ lie much above the configurations obtained via Methods I and II, which shows that it would require a much longer (numerically inaccessible) aging time to produce a simialr stability.}
Overall, this suggests that the annealing procedures I and II at small rates lead to more stable KA states with lower IS energies \rrev{for a modest computational cost.}   
 
\subsection{Fictive temperatures}

\label{fictive}

To further quantify the stability of the annealed states, we estimate their fictive temperature $T_f$. To this end, we fit the temperature evolution of the equilibrium IS energy of the KA model as $e_{IS}^{\rm eq}(T) = a/T + b$ (with $a$, $b$ fitting parameters), shown by the dashed blue line in Fig.~\ref{fig2}. This is a well-known temperature dependence~\cite{sastry2001relationship,elmatad2010corresponding}. 

We can then directly read-off the value of the fictive temperatures for the \rev{KA configurations obtained from direct aging or annealing} by the identification $e_{IS} = e_{IS}^{\rm eq}(T_f)$. This is shown with the \rev{black and} purple dashed-dotted lines, \rev{respectively} in Fig.~\ref{fig2}. We find that in a direct KA simulation, the lowest IS energies correspond to $T_f \simeq 0.386$, whereas the lowest IS energies for methods I and II give $T_f \simeq 0.355$. The latter is within $18\%$ of the experimental glass transition temperature estimated for this system, $T_g \simeq 0.30$. These \rev{fictive temperature} values confirm the enhanced stability of the annealed KA configurations. 

\subsection{Relaxation timescales} 

\begin{figure}
\includegraphics[width=8.5cm]{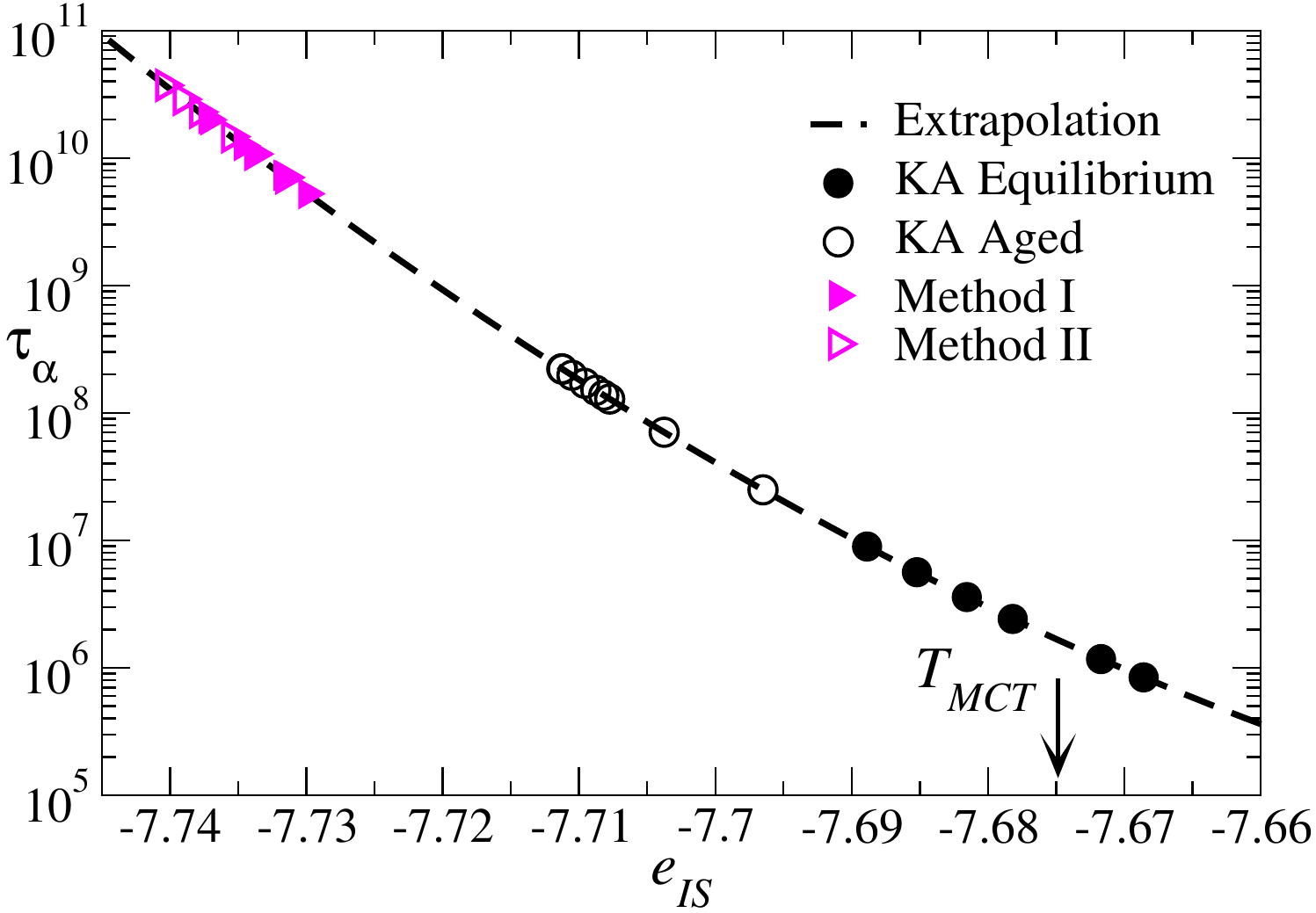}
\caption{Parametric plot of the estimated relaxation time $\tau_\alpha$ versus the average IS energy per particle $e_\mathrm{IS}$. The \rev{dashed} line combines the estimate of $\tau_\alpha(T)$ using a parabolic fit with an affine dependence of $e_\mathrm{IS}$ with $1/T$. We can then report the IS energies obtained by direct aging in the KA model, or by annealing the KA$_1$ model with methods I and II, and convert them into estimated relaxation times.}
\label{fig3}
\end{figure}

To determine a dynamic speedup gained by the annealing protocols devised above, we convert the obtained IS energies (or, equivalently, fictive temperatures) into an equilibrium relaxation timescale. To do this, we first need to extrapolate the equilibrium relaxation time $\tau_\alpha(T)$ of the KA model to lower temperatures, to infer relaxation timescales that are too large to be directly measured~\cite{berthier2020measure}. We use a parabolic fit of the temperature dependence of $\tau_\alpha(T)$~\cite{elmatad2010corresponding}\rev{, namely $\tau_{\alpha, {\rm p}}(T)= \tau_{0,{\rm p}}e^{J(1/T-1/T_{\rm p})^2}$ (with $\tau_{0,{\rm p}}$, $J$ and $T_{\rm p}$ fitting parameters). Consequently, for each value of the IS energy in Fig.~\ref{fig2}, we determine the corresponding fictive temperature $T_f$ (thanks to the method explained in the previous section) and we compute the extrapolated $\alpha$-relaxation time $\tau_\alpha=\tau_{\alpha,{\rm p}}(T_f)$ from the parabolic fit. In Fig.~\ref{fig3}, we display the parametric plot $\tau_\alpha(e_{IS})$ for IS energies obtained at various temperatures, either by direct aging in the KA model or by annealing KA$_1$ configurations with the slowest annealing rates.}
 The lowest IS energies obtained for the annealed configurations provide much larger estimates of the corresponding relaxation timescales, with a speedup factor of about $10^2 - 10^3$. Therefore, we conclude that the speedup factor obtained for the KA$_1$ model with $\delta =0.5 \%$ translates into a similar speedup for the original KA model, for an equivalent computational effort.  

We recall that this very large speedup factor is obtained keeping constant the total computational timescale involved in the preparation of the KA configurations. \rev{In other words, the speedup offered by the present algorithms are totally costless, unlike all other methods described in the introduction.} We did not attempt to combine our approach to any other technique, such as parallel tempering, graphic cards, or longer simulation times. This would provide even more stable configurations, at the expense of increased computational time and, for some of these methods, a different scaling of the efficiency with system size.  

\subsection{Rheology}

We next examine the stability of the annealed KA configurations against shear deformation. It has recently been shown that the stability of glassy configurations qualitatively affects the nature of the yielding transition, with a sharp ductile-to-brittle transition with increasing stability~\cite{ozawa2018random,ozawa2020role,singh2020brittle,yeh2020glass,bhaumik2019role}. This transition is characterised by the emergence, in large enough systems, of a macroscopic discontinuity in the stress-strain curves, accompanied by the formation of macroscopic failure taking the form of a system-spanning shear-band. \rev{Despite scores of rheological studies of the KA model, this transition has not been observed in this model so far.}

To study the rheology of stable KA configurations, we need to prepare larger configurations. We first produce very large KA / KA$_1$ samples by replicating $3^3$ systems of $N=1000$ / $1005$ particles to obtain larger samples of $N = 27000$ / $27135$ particles. These replicated systems are further aged for $10^6$ MC steps at temperature $T=0.36$. The KA$_{1}$ samples are then annealed to KA states using both methods I and II, with $t_\mathrm{MC}=5 \times 10^4$, $p_\omega=1/50$ and $d\omega=5\times 10^{-3}$. 

\begin{figure}
\includegraphics[width=8.5cm]{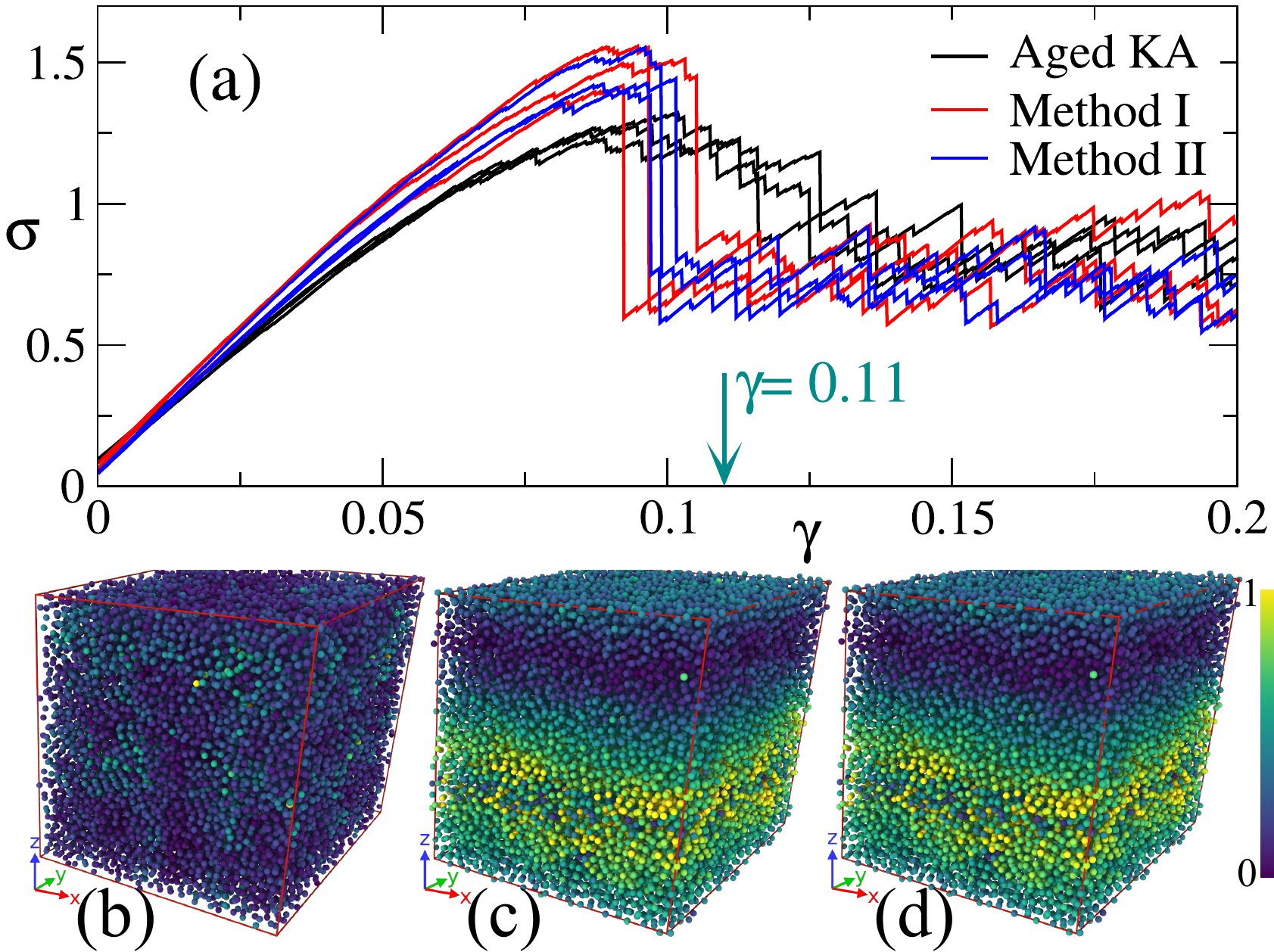}
\caption{(a) Stress-strain curves for aged KA and annealed samples with methods I and II. We report three independent loading curves for each case. 
The smooth stress overshoot of the aged KA turns into a sharp stress drop for the stable annealed samples. Snapshots of the non-affine displacement between $\gamma = 0$ and $\gamma = 0.11$ for (b) a KA sample, (c) an annealed sample with method I and (d) an annealed sample with method II. \rev{The color of the particles encodes the absolute magnitude of their non-affine displacement in units of $\sigma_{AA}$.}}
\label{fig4}
\end{figure}

Using these KA configurations, we perform a constant-volume athermal quasi-static shear protocol in the $xz$-plane with a strain increment $\Delta \gamma = 10^{-4}$ \rev{using  Lees-Edwards periodic boundary conditions. Each strain increment is followed by an energy minimization using the conjugate-gradient method.} In Fig.~\ref{fig4}(a), we present the stress-strain curves \rrev{$\sigma(\gamma)$} for three different samples for each of the three different preparation protocols (aged KA, methods I and II). In all cases, we observe an elastic regime, a weakening due to small plastic events, followed by a stress drop at the yielding transition, before reaching a steady-state regime at large deformation. For the aged KA samples, the yielding transition after the stress overshoot is the result of several plastic events, resulting in a modest stress drop and a relatively homogeneous strain field [see snapshot in Fig.~\ref{fig4}(b)]. The two annealing protocols provide KA samples with much lower fictive temperatures. This results in unique, sharp and macroscopic stress drops in the stress-strain curves of all samples, associated with system-spanning shear-bands that are formed within a single energy minimisation, and a highly heterogeneous plastic deformation field [see snapshots in Figs.~\ref{fig4}(c,d)]. The strong shear localisation at the yielding transition is correlated with the increased stability of the system~\cite{varnik2003shear,shi2005strain,ozawa2018random,singh2020brittle,yeh2020glass,ketkaew2018mechanical,bhaumik2019role,ozawa2020role,kapteijns2019fast}, which further confirms that the proposed annealing methods produce highly stable KA glass configurations.

\rrev{Physical information about the properties of stable KA configurations is also contained in the elastic regime at small deformation. In particular, the shear modulus $G$ describes the elastic response of the system: $G=\sigma/\gamma$ for small enough $\gamma$. For the determination of the shear modulus, we revert to system sizes $N=1000$ / $1005$ and we perform athermal quasi-static simulations for the equilibrium, aged, and annealed KA samples corresponding to $t_{MC}$= $7.5 \times 10^7$. The slope of the stress-strain curve in the range of $0$ to $1\%$ strain is plotted as a function of either the temperature (for equilibrium samples) or the fictive temperature (for non-equilibrium glasses) determined in Sec.~\ref{fictive}.}

\begin{figure}
\includegraphics[width=8.5cm]{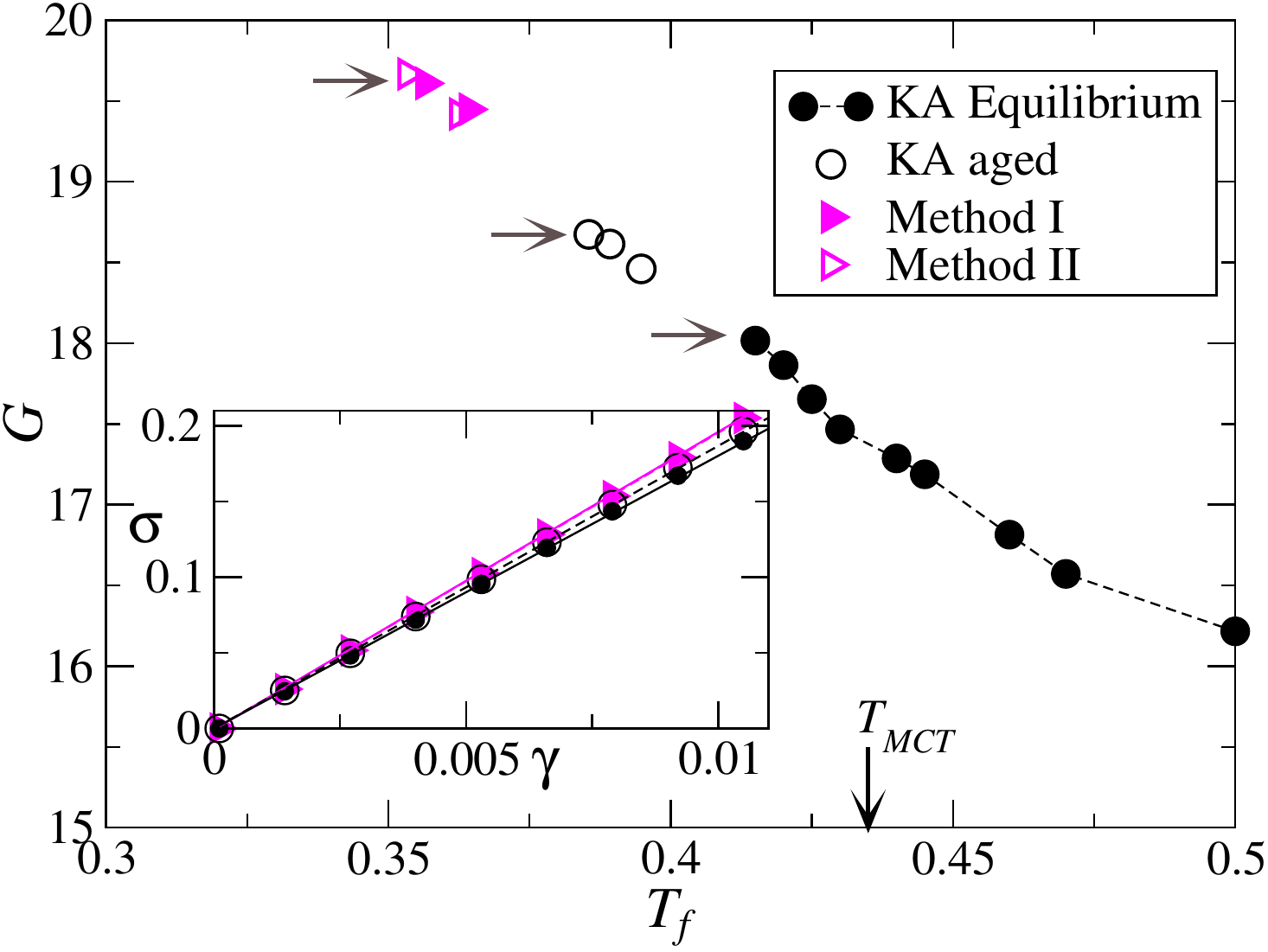}
\caption{\rrev {The shear modulus $G$ for various equilibrium, aged and annealed KA samples as a function of the fictive temperature (or the actual temperature for equilibrium samples). The annealed KA samples are achieved using methods I and II with $t_\mathrm{MC}=7.5\times10^{7}$. Inset: stress-strain curves in the small $\gamma$ regime used to determine $G$, for samples marked with horizontal arrows.}} 
\label{fig6}
\end{figure}

\rrev{The results are shown in Fig.~\ref{fig6}. The data for equilibrium configurations show that $G$ increases modestly over the simulated temperature regime (by about $10\%$). The aged KA configurations display a shear modulus that is about $2\%$ larger whereas the annealed glasses obtained using methods I and II exhibit an $8\%$ increase instead, confirming again that the annealed configurations are more stable than well-aged KA glasses. The reported trend is consistent with various earlier studies~\cite{mizuno2013measuring,wang2019low,shakerpoor2020stability,lerner2019mechanical}.}

\rev{These results demonstrate that brittle yielding can now be analysed in computer simulations of metallic glasses as well, and especially in the KA model which is one of the most studied models. In particular, we have demonstrated that the behaviour of KA glasses can turn from ductile to brittle by tuning the degree of annealing of the configuration before shearing, just like in experiments. This result was first shown in polydisperse soft sphere systems~\cite{ozawa2018random}, and now in the KA model which mimics metallic glasses with a small number of components. Our annealing procedure thus provides a systematic way of tuning the mechanical response of simulated metallic glasses and complement other preparation protocols, like cycling shear at finite temperature and finite shear rate~\cite{das2018annealing}. This new possibility also opens interesting research avenues to understand, for instance, the correlation between deformation in the brittle regime and different structural indicators like the ones studied in Ref.~\cite{richard2020predicting}, but also locally favoured structures, which are well documented in the KA model~\cite{coslovich2011locally,coslovich2016structure}. In addition, within the KA model, the influence of attractive forces on the yielding behaviour could also be investigated in computer simulations by direct comparison with a purely repulsive model (particles interacting via the WCA potential~\cite{andersen1971relationship}). These attractive forces are known to affect the equilibrium behaviour of supercooled liquids quantitatively~\cite{berthier2009nonperturbative} and experiments suggest that the rheology of attractive glasses is also quantitatively different from their repulsive counterparts~\cite{koumakis2011two}.}

\subsection{Calorimetric measurements}

We finally perform calorimetric measurements on systems of size $N=1000$ / $1005$ to study the stability of the generated glasses in the spirit of experiments performed on vapor-deposited ultrastable glasses~\cite{swallen2007organic,dawson2011highly}. Our goal is to monitor the onset temperature $T_\mathrm{on}$ at which the potential energy per particle $e(T)$ shows a brutal change of slope from its low-temperature glassy behavior when the glass sample is heated at constant rate. \rev{Note that this temperature is different from the onset temperature of glassy behaviour $T_0$ mentioned before, as $T_\mathrm{on}$ is a non-equilibrium, rate-dependent quantity while $T_0$ is measured at equilibrium.}

\begin{figure}
\includegraphics[width=8.5cm]{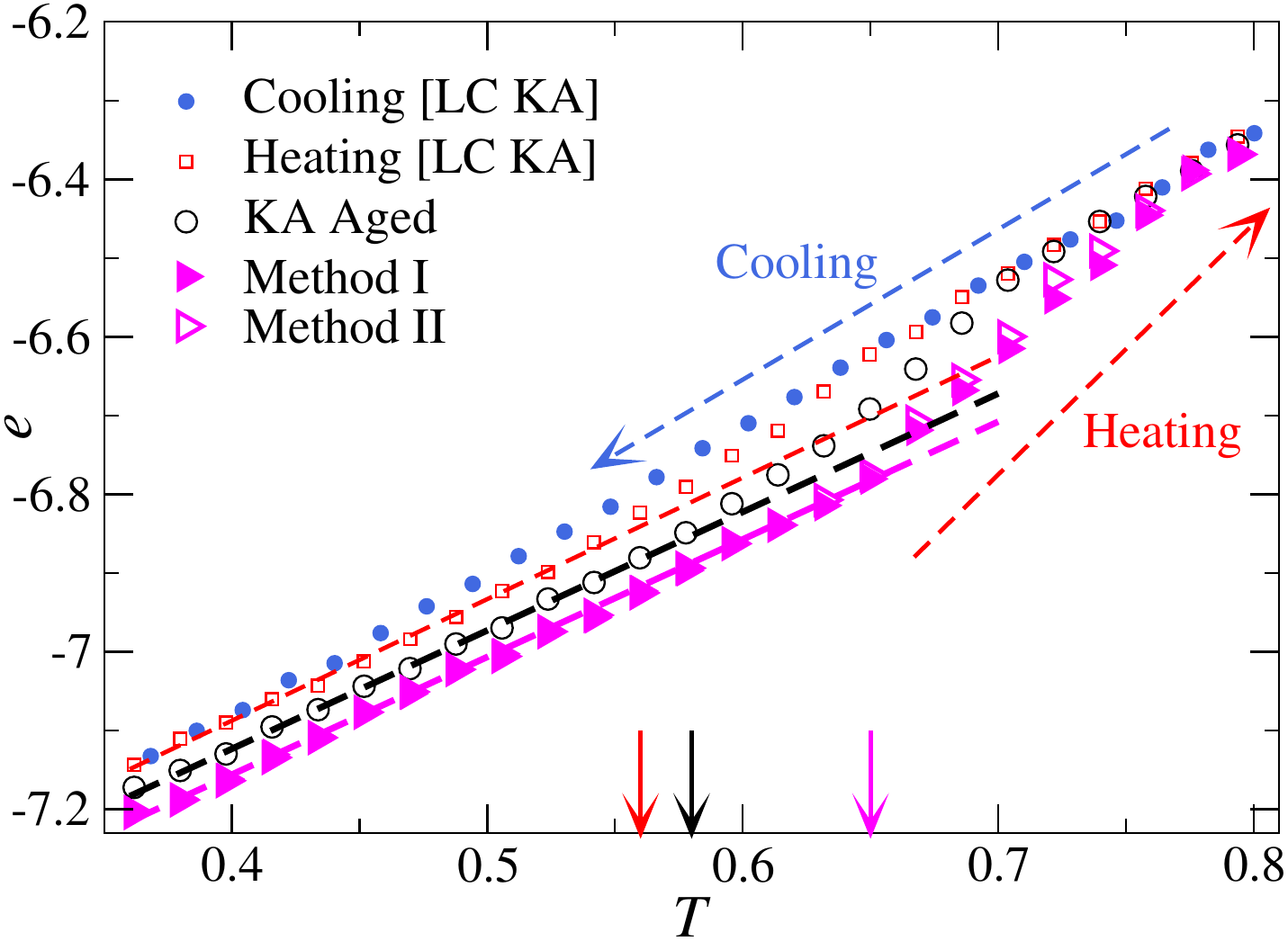}
\caption{Potential energy per particle $e$ of various glasses heated at constant rate $10^{-6}$: the liquid-cooled glass (LC KA) and the aged KA glass show relatively lower $T_{\mathrm on}$, compared to the glasses generated by the two annealing methods I and II. The onset temperatures \rev{$T_{\mathrm on}$} are marked by arrows. For the liquid-cooled glass cooled at constant rate $10^{-7}$, we also show the cooling curve.}
\label{fig5}
\end{figure}

In Fig.~\ref{fig5}, we compare four different glasses at the same heating rate of $10^{-6}$: (i) a glass prepared from an equilibrium configuration at $T=1.36$ cooled at a constant rate of $10^{-7}$; (ii) a KA sample aged at $T=0.36$ during $t_w=10^8$; (iii/iv) annealed samples prepared thanks to methods I/II at the same temperature $T=0.36$ and the lowest annealing rate (with $t_{MC}=7.5 \times 10^7$). For the liquid-cooled glass, we estimate the onset temperature $T_\mathrm{on}=0.56$. The well-aged KA sample shows a moderately larger onset temperature, $T_{\rm on}=0.58$, while the two annealed glasses display a higher $T_\mathrm{on}=0.65$, which again reflects the much larger kinetic stability reached using the annealing methods proposed in this work. 

\section{Conclusion}

\label{sec:conclusion}

To summarise, we have examined the possibility of using the speedup offered by the swap Monte Carlo algorithm in the \rev{extended Kob-Andersen (KA$_1$)} model to access \rev{low-energy} states in the \rev{original Kob-Andersen (KA)} model where swap moves are inefficient. We found that an equilibrium method introduces a bottleneck with equilibration times that are as large as in the original KA model and therefore this method does not provide any significant speedup. We have however introduced two non-equilibrium annealing methods that produce very stable glassy configurations of the KA model at equivalent computational cost, with a speedup of about 2-3 orders of magnitude. The achieved glass states have a significantly lower inherent structure energy compared to the ones obtained from direct aging in the KA model, they have lower fictive temperatures, and mechanical and calorimetric properties that indeed correspond to enhanced kinetic stabilities. 

We have thus developed a computationally cheap method to produce KA glassy configurations that are very stable. Unlike parallel tempering, transition path sampling, or ghost insertion method, the methods proposed here scale very well with system size, and are conceptually very simple. 
\rev{The present algorithm thus outperforms these more complicated algorithms.}
We believe that our strategy is  generic, and it can be implemented in other glass-formers with a small number of components. We also believe that combining the annealing methods with a parallel tempering scheme or graphic card simulations would allow the production of even more stable systems. These would prove useful for further investigations of physical properties of highly stable metallic glasses\rev{ using the well-studied KA model.}
 
\begin{acknowledgments}
We thank C. Cammarota for forcing us to think more deeply about this problem, and D. Coslovich and M. Ozawa for useful exchanges. This work was supported by a grant from the Simons Foundation (Grant No. 454933, L. B.). B. G. acknowledges support by Capital Fund Management - Fondation pour la Recherche. The data that support the findings of this study are available from the corresponding author upon reasonable request.
\end{acknowledgments}

\section*{DATA AVAILABILITY}

The data that support the findings of this study are available
from the corresponding author upon reasonable request.

\bibliography{metallic-glasses.bib}{}

\begin{thebibliography}{55}%
\makeatletter
\providecommand \@ifxundefined [1]{%
 \@ifx{#1\undefined}
}%
\providecommand \@ifnum [1]{%
 \ifnum #1\expandafter \@firstoftwo
 \else \expandafter \@secondoftwo
 \fi
}%
\providecommand \@ifx [1]{%
 \ifx #1\expandafter \@firstoftwo
 \else \expandafter \@secondoftwo
 \fi
}%
\providecommand \natexlab [1]{#1}%
\providecommand \enquote  [1]{``#1''}%
\providecommand \bibnamefont  [1]{#1}%
\providecommand \bibfnamefont [1]{#1}%
\providecommand \citenamefont [1]{#1}%
\providecommand \href@noop [0]{\@secondoftwo}%
\providecommand \href [0]{\begingroup \@sanitize@url \@href}%
\providecommand \@href[1]{\@@startlink{#1}\@@href}%
\providecommand \@@href[1]{\endgroup#1\@@endlink}%
\providecommand \@sanitize@url [0]{\catcode `\\12\catcode `\$12\catcode
  `\&12\catcode `\#12\catcode `\^12\catcode `\_12\catcode `\%12\relax}%
\providecommand \@@startlink[1]{}%
\providecommand \@@endlink[0]{}%
\providecommand \url  [0]{\begingroup\@sanitize@url \@url }%
\providecommand \@url [1]{\endgroup\@href {#1}{\urlprefix }}%
\providecommand \urlprefix  [0]{URL }%
\providecommand \Eprint [0]{\href }%
\providecommand \doibase [0]{http://dx.doi.org/}%
\providecommand \selectlanguage [0]{\@gobble}%
\providecommand \bibinfo  [0]{\@secondoftwo}%
\providecommand \bibfield  [0]{\@secondoftwo}%
\providecommand \translation [1]{[#1]}%
\providecommand \BibitemOpen [0]{}%
\providecommand \bibitemStop [0]{}%
\providecommand \bibitemNoStop [0]{.\EOS\space}%
\providecommand \EOS [0]{\spacefactor3000\relax}%
\providecommand \BibitemShut  [1]{\csname bibitem#1\endcsname}%
\let\auto@bib@innerbib\@empty
\bibitem [{\citenamefont {Berthier}\ and\ \citenamefont
  {Biroli}(2011)}]{berthier2011theoretical}%
  \BibitemOpen
  \bibfield  {author} {\bibinfo {author} {\bibfnamefont {L.}~\bibnamefont
  {Berthier}}\ and\ \bibinfo {author} {\bibfnamefont {G.}~\bibnamefont
  {Biroli}},\ }\href@noop {} {\bibfield  {journal} {\bibinfo  {journal} {Rev.
  Mod. Phys}\ }\textbf {\bibinfo {volume} {83}},\ \bibinfo {pages} {587}
  (\bibinfo {year} {2011})}\BibitemShut {NoStop}%
\bibitem [{\citenamefont {Berthier}\ \emph {et~al.}(2016)\citenamefont
  {Berthier}, \citenamefont {Coslovich}, \citenamefont {Ninarello},\ and\
  \citenamefont {Ozawa}}]{berthier2016equilibrium}%
  \BibitemOpen
  \bibfield  {author} {\bibinfo {author} {\bibfnamefont {L.}~\bibnamefont
  {Berthier}}, \bibinfo {author} {\bibfnamefont {D.}~\bibnamefont {Coslovich}},
  \bibinfo {author} {\bibfnamefont {A.}~\bibnamefont {Ninarello}}, \ and\
  \bibinfo {author} {\bibfnamefont {M.}~\bibnamefont {Ozawa}},\ }\href@noop {}
  {\bibfield  {journal} {\bibinfo  {journal} {Phys. Rev. Lett}\ }\textbf
  {\bibinfo {volume} {116}},\ \bibinfo {pages} {238002} (\bibinfo {year}
  {2016})}\BibitemShut {NoStop}%
\bibitem [{\citenamefont {Ninarello}\ \emph {et~al.}(2017)\citenamefont
  {Ninarello}, \citenamefont {Berthier},\ and\ \citenamefont
  {Coslovich}}]{ninarello2017models}%
  \BibitemOpen
  \bibfield  {author} {\bibinfo {author} {\bibfnamefont {A.}~\bibnamefont
  {Ninarello}}, \bibinfo {author} {\bibfnamefont {L.}~\bibnamefont {Berthier}},
  \ and\ \bibinfo {author} {\bibfnamefont {D.}~\bibnamefont {Coslovich}},\
  }\href@noop {} {\bibfield  {journal} {\bibinfo  {journal} {Phys. Rev. X}\
  }\textbf {\bibinfo {volume} {7}},\ \bibinfo {pages} {021039} (\bibinfo {year}
  {2017})}\BibitemShut {NoStop}%
\bibitem [{\citenamefont {Berthier}\ \emph
  {et~al.}(2019{\natexlab{a}})\citenamefont {Berthier}, \citenamefont
  {Flenner}, \citenamefont {Fullerton}, \citenamefont {Scalliet},\ and\
  \citenamefont {Singh}}]{berthier2019efficient}%
  \BibitemOpen
  \bibfield  {author} {\bibinfo {author} {\bibfnamefont {L.}~\bibnamefont
  {Berthier}}, \bibinfo {author} {\bibfnamefont {E.}~\bibnamefont {Flenner}},
  \bibinfo {author} {\bibfnamefont {C.~J.}\ \bibnamefont {Fullerton}}, \bibinfo
  {author} {\bibfnamefont {C.}~\bibnamefont {Scalliet}}, \ and\ \bibinfo
  {author} {\bibfnamefont {M.}~\bibnamefont {Singh}},\ }\href@noop {}
  {\bibfield  {journal} {\bibinfo  {journal} {Journal of Statistical Mechanics:
  Theory and Experiment}\ }\textbf {\bibinfo {volume} {2019}},\ \bibinfo
  {pages} {064004} (\bibinfo {year} {2019}{\natexlab{a}})}\BibitemShut
  {NoStop}%
\bibitem [{\citenamefont {Fullerton}\ and\ \citenamefont
  {Berthier}(2017)}]{fullerton2017density}%
  \BibitemOpen
  \bibfield  {author} {\bibinfo {author} {\bibfnamefont {C.~J.}\ \bibnamefont
  {Fullerton}}\ and\ \bibinfo {author} {\bibfnamefont {L.}~\bibnamefont
  {Berthier}},\ }\href@noop {} {\bibfield  {journal} {\bibinfo  {journal} {EPL
  (Europhysics Letters)}\ }\textbf {\bibinfo {volume} {119}},\ \bibinfo {pages}
  {36003} (\bibinfo {year} {2017})}\BibitemShut {NoStop}%
\bibitem [{\citenamefont {Berthier}\ \emph {et~al.}(2017)\citenamefont
  {Berthier}, \citenamefont {Charbonneau}, \citenamefont {Coslovich},
  \citenamefont {Ninarello}, \citenamefont {Ozawa},\ and\ \citenamefont
  {Yaida}}]{berthier2017configurational}%
  \BibitemOpen
  \bibfield  {author} {\bibinfo {author} {\bibfnamefont {L.}~\bibnamefont
  {Berthier}}, \bibinfo {author} {\bibfnamefont {P.}~\bibnamefont
  {Charbonneau}}, \bibinfo {author} {\bibfnamefont {D.}~\bibnamefont
  {Coslovich}}, \bibinfo {author} {\bibfnamefont {A.}~\bibnamefont
  {Ninarello}}, \bibinfo {author} {\bibfnamefont {M.}~\bibnamefont {Ozawa}}, \
  and\ \bibinfo {author} {\bibfnamefont {S.}~\bibnamefont {Yaida}},\
  }\href@noop {} {\bibfield  {journal} {\bibinfo  {journal} {PNAS}\ }\textbf
  {\bibinfo {volume} {114}},\ \bibinfo {pages} {11356} (\bibinfo {year}
  {2017})}\BibitemShut {NoStop}%
\bibitem [{\citenamefont {Berthier}\ \emph
  {et~al.}(2019{\natexlab{b}})\citenamefont {Berthier}, \citenamefont
  {Charbonneau}, \citenamefont {Ninarello}, \citenamefont {Ozawa},\ and\
  \citenamefont {Yaida}}]{berthier2019zero}%
  \BibitemOpen
  \bibfield  {author} {\bibinfo {author} {\bibfnamefont {L.}~\bibnamefont
  {Berthier}}, \bibinfo {author} {\bibfnamefont {P.}~\bibnamefont
  {Charbonneau}}, \bibinfo {author} {\bibfnamefont {A.}~\bibnamefont
  {Ninarello}}, \bibinfo {author} {\bibfnamefont {M.}~\bibnamefont {Ozawa}}, \
  and\ \bibinfo {author} {\bibfnamefont {S.}~\bibnamefont {Yaida}},\
  }\href@noop {} {\bibfield  {journal} {\bibinfo  {journal} {Nat. Commun.}\
  }\textbf {\bibinfo {volume} {10}},\ \bibinfo {pages} {1} (\bibinfo {year}
  {2019}{\natexlab{b}})}\BibitemShut {NoStop}%
\bibitem [{\citenamefont {Scalliet}\ \emph {et~al.}(2019)\citenamefont
  {Scalliet}, \citenamefont {Berthier},\ and\ \citenamefont
  {Zamponi}}]{scalliet2019nature}%
  \BibitemOpen
  \bibfield  {author} {\bibinfo {author} {\bibfnamefont {C.}~\bibnamefont
  {Scalliet}}, \bibinfo {author} {\bibfnamefont {L.}~\bibnamefont {Berthier}},
  \ and\ \bibinfo {author} {\bibfnamefont {F.}~\bibnamefont {Zamponi}},\
  }\href@noop {} {\bibfield  {journal} {\bibinfo  {journal} {Nature
  communications}\ }\textbf {\bibinfo {volume} {10}},\ \bibinfo {pages} {1}
  (\bibinfo {year} {2019})}\BibitemShut {NoStop}%
\bibitem [{\citenamefont {Wang}\ \emph {et~al.}(2019)\citenamefont {Wang},
  \citenamefont {Ninarello}, \citenamefont {Guan}, \citenamefont {Berthier},
  \citenamefont {Szamel},\ and\ \citenamefont {Flenner}}]{wang2019low}%
  \BibitemOpen
  \bibfield  {author} {\bibinfo {author} {\bibfnamefont {L.}~\bibnamefont
  {Wang}}, \bibinfo {author} {\bibfnamefont {A.}~\bibnamefont {Ninarello}},
  \bibinfo {author} {\bibfnamefont {P.}~\bibnamefont {Guan}}, \bibinfo {author}
  {\bibfnamefont {L.}~\bibnamefont {Berthier}}, \bibinfo {author}
  {\bibfnamefont {G.}~\bibnamefont {Szamel}}, \ and\ \bibinfo {author}
  {\bibfnamefont {E.}~\bibnamefont {Flenner}},\ }\href@noop {} {\bibfield
  {journal} {\bibinfo  {journal} {Nature communications}\ }\textbf {\bibinfo
  {volume} {10}},\ \bibinfo {pages} {1} (\bibinfo {year} {2019})}\BibitemShut
  {NoStop}%
\bibitem [{\citenamefont {Ozawa}\ \emph {et~al.}(2019)\citenamefont {Ozawa},
  \citenamefont {Scalliet}, \citenamefont {Ninarello},\ and\ \citenamefont
  {Berthier}}]{ozawa2019does}%
  \BibitemOpen
  \bibfield  {author} {\bibinfo {author} {\bibfnamefont {M.}~\bibnamefont
  {Ozawa}}, \bibinfo {author} {\bibfnamefont {C.}~\bibnamefont {Scalliet}},
  \bibinfo {author} {\bibfnamefont {A.}~\bibnamefont {Ninarello}}, \ and\
  \bibinfo {author} {\bibfnamefont {L.}~\bibnamefont {Berthier}},\ }\href@noop
  {} {\bibfield  {journal} {\bibinfo  {journal} {J. Chem. Phys.}\ }\textbf
  {\bibinfo {volume} {151}},\ \bibinfo {pages} {084504} (\bibinfo {year}
  {2019})}\BibitemShut {NoStop}%
\bibitem [{\citenamefont {Berthier}\ \emph
  {et~al.}(2019{\natexlab{c}})\citenamefont {Berthier}, \citenamefont
  {Charbonneau},\ and\ \citenamefont {Kundu}}]{berthier2019finite}%
  \BibitemOpen
  \bibfield  {author} {\bibinfo {author} {\bibfnamefont {L.}~\bibnamefont
  {Berthier}}, \bibinfo {author} {\bibfnamefont {P.}~\bibnamefont
  {Charbonneau}}, \ and\ \bibinfo {author} {\bibfnamefont {J.}~\bibnamefont
  {Kundu}},\ }\href@noop {} {\bibfield  {journal} {\bibinfo  {journal} {arXiv
  preprint arXiv:1912.11510}\ } (\bibinfo {year}
  {2019}{\natexlab{c}})}\BibitemShut {NoStop}%
\bibitem [{\citenamefont {Khomenko}\ \emph {et~al.}(2020)\citenamefont
  {Khomenko}, \citenamefont {Scalliet}, \citenamefont {Berthier}, \citenamefont
  {Reichman},\ and\ \citenamefont {Zamponi}}]{khomenko2020depletion}%
  \BibitemOpen
  \bibfield  {author} {\bibinfo {author} {\bibfnamefont {D.}~\bibnamefont
  {Khomenko}}, \bibinfo {author} {\bibfnamefont {C.}~\bibnamefont {Scalliet}},
  \bibinfo {author} {\bibfnamefont {L.}~\bibnamefont {Berthier}}, \bibinfo
  {author} {\bibfnamefont {D.~R.}\ \bibnamefont {Reichman}}, \ and\ \bibinfo
  {author} {\bibfnamefont {F.}~\bibnamefont {Zamponi}},\ }\href@noop {}
  {\bibfield  {journal} {\bibinfo  {journal} {Physical Review Letters}\
  }\textbf {\bibinfo {volume} {124}},\ \bibinfo {pages} {225901} (\bibinfo
  {year} {2020})}\BibitemShut {NoStop}%
\bibitem [{\citenamefont {Guiselin}\ \emph {et~al.}(2020)\citenamefont
  {Guiselin}, \citenamefont {Berthier},\ and\ \citenamefont
  {Tarjus}}]{guiselin2020random}%
  \BibitemOpen
  \bibfield  {author} {\bibinfo {author} {\bibfnamefont {B.}~\bibnamefont
  {Guiselin}}, \bibinfo {author} {\bibfnamefont {L.}~\bibnamefont {Berthier}},
  \ and\ \bibinfo {author} {\bibfnamefont {G.}~\bibnamefont {Tarjus}},\
  }\href@noop {} {\bibfield  {journal} {\bibinfo  {journal} {arXiv preprint
  arXiv:2004.10555}\ } (\bibinfo {year} {2020})}\BibitemShut {NoStop}%
\bibitem [{\citenamefont {Kob}\ and\ \citenamefont
  {Andersen}(1995)}]{kob1995testing}%
  \BibitemOpen
  \bibfield  {author} {\bibinfo {author} {\bibfnamefont {W.}~\bibnamefont
  {Kob}}\ and\ \bibinfo {author} {\bibfnamefont {H.~C.}\ \bibnamefont
  {Andersen}},\ }\href@noop {} {\bibfield  {journal} {\bibinfo  {journal}
  {Phys. Rev. E}\ }\textbf {\bibinfo {volume} {51}},\ \bibinfo {pages} {4626}
  (\bibinfo {year} {1995})}\BibitemShut {NoStop}%
\bibitem [{\citenamefont {Flenner}\ and\ \citenamefont
  {Szamel}(2006)}]{flenner2006hybrid}%
  \BibitemOpen
  \bibfield  {author} {\bibinfo {author} {\bibfnamefont {E.}~\bibnamefont
  {Flenner}}\ and\ \bibinfo {author} {\bibfnamefont {G.}~\bibnamefont
  {Szamel}},\ }\href@noop {} {\bibfield  {journal} {\bibinfo  {journal} {Phys.
  Rev. E}\ }\textbf {\bibinfo {volume} {73}},\ \bibinfo {pages} {061505}
  (\bibinfo {year} {2006})}\BibitemShut {NoStop}%
\bibitem [{\citenamefont {Coslovich}\ \emph {et~al.}(2018)\citenamefont
  {Coslovich}, \citenamefont {Ozawa},\ and\ \citenamefont
  {Kob}}]{coslovich2018dynamic}%
  \BibitemOpen
  \bibfield  {author} {\bibinfo {author} {\bibfnamefont {D.}~\bibnamefont
  {Coslovich}}, \bibinfo {author} {\bibfnamefont {M.}~\bibnamefont {Ozawa}}, \
  and\ \bibinfo {author} {\bibfnamefont {W.}~\bibnamefont {Kob}},\ }\href@noop
  {} {\bibfield  {journal} {\bibinfo  {journal} {Eur. Phys. J. E}\ }\textbf
  {\bibinfo {volume} {41}},\ \bibinfo {pages} {62} (\bibinfo {year}
  {2018})}\BibitemShut {NoStop}%
\bibitem [{\citenamefont {Schr{\o}der}\ and\ \citenamefont
  {Dyre}(2020)}]{schroder2020solid}%
  \BibitemOpen
  \bibfield  {author} {\bibinfo {author} {\bibfnamefont {T.~B.}\ \bibnamefont
  {Schr{\o}der}}\ and\ \bibinfo {author} {\bibfnamefont {J.~C.}\ \bibnamefont
  {Dyre}},\ }\href@noop {} {\bibfield  {journal} {\bibinfo  {journal} {The
  Journal of Chemical Physics}\ }\textbf {\bibinfo {volume} {152}},\ \bibinfo
  {pages} {141101} (\bibinfo {year} {2020})}\BibitemShut {NoStop}%
\bibitem [{\citenamefont {Turci}\ \emph {et~al.}(2017)\citenamefont {Turci},
  \citenamefont {Royall},\ and\ \citenamefont
  {Speck}}]{turci2017nonequilibrium}%
  \BibitemOpen
  \bibfield  {author} {\bibinfo {author} {\bibfnamefont {F.}~\bibnamefont
  {Turci}}, \bibinfo {author} {\bibfnamefont {C.~P.}\ \bibnamefont {Royall}}, \
  and\ \bibinfo {author} {\bibfnamefont {T.}~\bibnamefont {Speck}},\
  }\href@noop {} {\bibfield  {journal} {\bibinfo  {journal} {Phys. Rev. X}\
  }\textbf {\bibinfo {volume} {7}},\ \bibinfo {pages} {031028} (\bibinfo {year}
  {2017})}\BibitemShut {NoStop}%
\bibitem [{\citenamefont {Faller}\ and\ \citenamefont
  {de~Pablo}(2003)}]{faller2003density}%
  \BibitemOpen
  \bibfield  {author} {\bibinfo {author} {\bibfnamefont {R.}~\bibnamefont
  {Faller}}\ and\ \bibinfo {author} {\bibfnamefont {J.~J.}\ \bibnamefont
  {de~Pablo}},\ }\href@noop {} {\bibfield  {journal} {\bibinfo  {journal} {J.
  Chem. Phys.}\ }\textbf {\bibinfo {volume} {119}},\ \bibinfo {pages} {4405}
  (\bibinfo {year} {2003})}\BibitemShut {NoStop}%
\bibitem [{\citenamefont {Jack}\ \emph {et~al.}(2011)\citenamefont {Jack},
  \citenamefont {Hedges}, \citenamefont {Garrahan},\ and\ \citenamefont
  {Chandler}}]{jack2011preparation}%
  \BibitemOpen
  \bibfield  {author} {\bibinfo {author} {\bibfnamefont {R.~L.}\ \bibnamefont
  {Jack}}, \bibinfo {author} {\bibfnamefont {L.~O.}\ \bibnamefont {Hedges}},
  \bibinfo {author} {\bibfnamefont {J.~P.}\ \bibnamefont {Garrahan}}, \ and\
  \bibinfo {author} {\bibfnamefont {D.}~\bibnamefont {Chandler}},\ }\href@noop
  {} {\bibfield  {journal} {\bibinfo  {journal} {Physical review letters}\
  }\textbf {\bibinfo {volume} {107}},\ \bibinfo {pages} {275702} (\bibinfo
  {year} {2011})}\BibitemShut {NoStop}%
\bibitem [{\citenamefont {Lyubimov}\ \emph {et~al.}(2013)\citenamefont
  {Lyubimov}, \citenamefont {Ediger},\ and\ \citenamefont
  {de~Pablo}}]{lyubimov2013model}%
  \BibitemOpen
  \bibfield  {author} {\bibinfo {author} {\bibfnamefont {I.}~\bibnamefont
  {Lyubimov}}, \bibinfo {author} {\bibfnamefont {M.~D.}\ \bibnamefont
  {Ediger}}, \ and\ \bibinfo {author} {\bibfnamefont {J.~J.}\ \bibnamefont
  {de~Pablo}},\ }\href@noop {} {\bibfield  {journal} {\bibinfo  {journal} {The
  Journal of chemical physics}\ }\textbf {\bibinfo {volume} {139}},\ \bibinfo
  {pages} {144505} (\bibinfo {year} {2013})}\BibitemShut {NoStop}%
\bibitem [{\citenamefont {Das}\ \emph {et~al.}(2018)\citenamefont {Das},
  \citenamefont {Parmar},\ and\ \citenamefont {Sastry}}]{das2018annealing}%
  \BibitemOpen
  \bibfield  {author} {\bibinfo {author} {\bibfnamefont {P.}~\bibnamefont
  {Das}}, \bibinfo {author} {\bibfnamefont {A.~D.}\ \bibnamefont {Parmar}}, \
  and\ \bibinfo {author} {\bibfnamefont {S.}~\bibnamefont {Sastry}},\
  }\href@noop {} {\bibfield  {journal} {\bibinfo  {journal} {arXiv preprint
  arXiv:1805.12476}\ } (\bibinfo {year} {2018})}\BibitemShut {NoStop}%
\bibitem [{\citenamefont {Priezjev}(2018)}]{priezjev2018molecular}%
  \BibitemOpen
  \bibfield  {author} {\bibinfo {author} {\bibfnamefont {N.~V.}\ \bibnamefont
  {Priezjev}},\ }\href@noop {} {\bibfield  {journal} {\bibinfo  {journal}
  {Journal of Non-Crystalline Solids}\ }\textbf {\bibinfo {volume} {479}},\
  \bibinfo {pages} {42} (\bibinfo {year} {2018})}\BibitemShut {NoStop}%
\bibitem [{\citenamefont {Bhowmik}\ \emph {et~al.}(2020)\citenamefont
  {Bhowmik}, \citenamefont {Iylin},\ and\ \citenamefont
  {Procaccia}}]{bhowmik2020thermodynamic}%
  \BibitemOpen
  \bibfield  {author} {\bibinfo {author} {\bibfnamefont {B.~P.}\ \bibnamefont
  {Bhowmik}}, \bibinfo {author} {\bibfnamefont {V.}~\bibnamefont {Iylin}}, \
  and\ \bibinfo {author} {\bibfnamefont {I.}~\bibnamefont {Procaccia}},\
  }\href@noop {} {\bibfield  {journal} {\bibinfo  {journal} {arXiv preprint
  arXiv:2003.14190}\ } (\bibinfo {year} {2020})}\BibitemShut {NoStop}%
\bibitem [{\citenamefont {Parmar}\ \emph {et~al.}(2020)\citenamefont {Parmar},
  \citenamefont {Ozawa},\ and\ \citenamefont
  {Berthier}}]{parmar2020ultrastable}%
  \BibitemOpen
  \bibfield  {author} {\bibinfo {author} {\bibfnamefont {A.~D.~S.}\
  \bibnamefont {Parmar}}, \bibinfo {author} {\bibfnamefont {M.}~\bibnamefont
  {Ozawa}}, \ and\ \bibinfo {author} {\bibfnamefont {L.}~\bibnamefont
  {Berthier}},\ }\href@noop {} {\bibfield  {journal} {\bibinfo  {journal}
  {arXiv preprint arXiv:2002.01317}\ } (\bibinfo {year} {2020})}\BibitemShut
  {NoStop}%
\bibitem [{\citenamefont {Berthier}\ and\ \citenamefont
  {Kob}(2007)}]{berthier2007monte}%
  \BibitemOpen
  \bibfield  {author} {\bibinfo {author} {\bibfnamefont {L.}~\bibnamefont
  {Berthier}}\ and\ \bibinfo {author} {\bibfnamefont {W.}~\bibnamefont {Kob}},\
  }\href@noop {} {\bibfield  {journal} {\bibinfo  {journal} {J. Phys. Condens.
  Matter}\ }\textbf {\bibinfo {volume} {19}},\ \bibinfo {pages} {205130}
  (\bibinfo {year} {2007})}\BibitemShut {NoStop}%
\bibitem [{\citenamefont {Newman}\ and\ \citenamefont
  {Barkema}(1999)}]{newman1999monte}%
  \BibitemOpen
  \bibfield  {author} {\bibinfo {author} {\bibfnamefont {M.}~\bibnamefont
  {Newman}}\ and\ \bibinfo {author} {\bibfnamefont {G.}~\bibnamefont
  {Barkema}},\ }\href@noop {} {\emph {\bibinfo {title} {Monte {C}arlo methods
  in statistical physics}}}\ (\bibinfo  {publisher} {Oxford University Press:
  New York, USA},\ \bibinfo {year} {1999})\BibitemShut {NoStop}%
\bibitem [{\citenamefont {Frenkel}\ and\ \citenamefont
  {Smit}(2001)}]{frenkel2001understanding}%
  \BibitemOpen
  \bibfield  {author} {\bibinfo {author} {\bibfnamefont {D.}~\bibnamefont
  {Frenkel}}\ and\ \bibinfo {author} {\bibfnamefont {B.}~\bibnamefont {Smit}},\
  }\href@noop {} {\emph {\bibinfo {title} {Understanding molecular simulation:
  from algorithms to applications}}},\ Vol.~\bibinfo {volume} {1}\ (\bibinfo
  {publisher} {Elsevier},\ \bibinfo {year} {2001})\BibitemShut {NoStop}%
\bibitem [{\citenamefont {Torrie}\ and\ \citenamefont
  {Valleau}(1977)}]{torrie1977monte}%
  \BibitemOpen
  \bibfield  {author} {\bibinfo {author} {\bibfnamefont {G.}~\bibnamefont
  {Torrie}}\ and\ \bibinfo {author} {\bibfnamefont {J.}~\bibnamefont
  {Valleau}},\ }\href@noop {} {\bibfield  {journal} {\bibinfo  {journal} {The
  Journal of chemical physics}\ }\textbf {\bibinfo {volume} {66}},\ \bibinfo
  {pages} {1402} (\bibinfo {year} {1977})}\BibitemShut {NoStop}%
\bibitem [{\citenamefont {Ferrenberg}\ and\ \citenamefont
  {Swendsen}(1988)}]{ferrenberg1988new}%
  \BibitemOpen
  \bibfield  {author} {\bibinfo {author} {\bibfnamefont {A.~M.}\ \bibnamefont
  {Ferrenberg}}\ and\ \bibinfo {author} {\bibfnamefont {R.~H.}\ \bibnamefont
  {Swendsen}},\ }\href@noop {} {\bibfield  {journal} {\bibinfo  {journal}
  {Physical review letters}\ }\textbf {\bibinfo {volume} {61}},\ \bibinfo
  {pages} {2635} (\bibinfo {year} {1988})}\BibitemShut {NoStop}%
\bibitem [{\citenamefont {Hartmann}(2011)}]{hartmann2011large}%
  \BibitemOpen
  \bibfield  {author} {\bibinfo {author} {\bibfnamefont {A.~K.}\ \bibnamefont
  {Hartmann}},\ }\href@noop {} {\bibfield  {journal} {\bibinfo  {journal} {The
  European Physical Journal B}\ }\textbf {\bibinfo {volume} {84}},\ \bibinfo
  {pages} {627} (\bibinfo {year} {2011})}\BibitemShut {NoStop}%
\bibitem [{\citenamefont {Van~Laarhoven}\ and\ \citenamefont
  {Aarts}(1987)}]{van1987simulated}%
  \BibitemOpen
  \bibfield  {author} {\bibinfo {author} {\bibfnamefont {P.~J.}\ \bibnamefont
  {Van~Laarhoven}}\ and\ \bibinfo {author} {\bibfnamefont {E.~H.}\ \bibnamefont
  {Aarts}},\ }in\ \href@noop {} {\emph {\bibinfo {booktitle} {Simulated
  annealing: Theory and applications}}}\ (\bibinfo  {publisher} {Springer},\
  \bibinfo {year} {1987})\ pp.\ \bibinfo {pages} {7--15}\BibitemShut {NoStop}%
\bibitem [{\citenamefont {Sastry}(2001)}]{sastry2001relationship}%
  \BibitemOpen
  \bibfield  {author} {\bibinfo {author} {\bibfnamefont {S.}~\bibnamefont
  {Sastry}},\ }\href@noop {} {\bibfield  {journal} {\bibinfo  {journal}
  {Nature}\ }\textbf {\bibinfo {volume} {409}},\ \bibinfo {pages} {164}
  (\bibinfo {year} {2001})}\BibitemShut {NoStop}%
\bibitem [{\citenamefont {Elmatad}\ \emph {et~al.}(2010)\citenamefont
  {Elmatad}, \citenamefont {Chandler},\ and\ \citenamefont
  {Garrahan}}]{elmatad2010corresponding}%
  \BibitemOpen
  \bibfield  {author} {\bibinfo {author} {\bibfnamefont {Y.~S.}\ \bibnamefont
  {Elmatad}}, \bibinfo {author} {\bibfnamefont {D.}~\bibnamefont {Chandler}}, \
  and\ \bibinfo {author} {\bibfnamefont {J.~P.}\ \bibnamefont {Garrahan}},\
  }\href@noop {} {\bibfield  {journal} {\bibinfo  {journal} {J. Phys. Chem. B}\
  }\textbf {\bibinfo {volume} {114}},\ \bibinfo {pages} {17113} (\bibinfo
  {year} {2010})}\BibitemShut {NoStop}%
\bibitem [{\citenamefont {Berthier}\ and\ \citenamefont
  {Ediger}(2020)}]{berthier2020measure}%
  \BibitemOpen
  \bibfield  {author} {\bibinfo {author} {\bibfnamefont {L.}~\bibnamefont
  {Berthier}}\ and\ \bibinfo {author} {\bibfnamefont {M.~D.}\ \bibnamefont
  {Ediger}},\ }\href@noop {} {\bibfield  {journal} {\bibinfo  {journal} {The
  Journal of Chemical Physics}\ }\textbf {\bibinfo {volume} {153}},\ \bibinfo
  {pages} {044501} (\bibinfo {year} {2020})}\BibitemShut {NoStop}%
\bibitem [{\citenamefont {Ozawa}\ \emph {et~al.}(2018)\citenamefont {Ozawa},
  \citenamefont {Berthier}, \citenamefont {Biroli}, \citenamefont {Rosso},\
  and\ \citenamefont {Tarjus}}]{ozawa2018random}%
  \BibitemOpen
  \bibfield  {author} {\bibinfo {author} {\bibfnamefont {M.}~\bibnamefont
  {Ozawa}}, \bibinfo {author} {\bibfnamefont {L.}~\bibnamefont {Berthier}},
  \bibinfo {author} {\bibfnamefont {G.}~\bibnamefont {Biroli}}, \bibinfo
  {author} {\bibfnamefont {A.}~\bibnamefont {Rosso}}, \ and\ \bibinfo {author}
  {\bibfnamefont {G.}~\bibnamefont {Tarjus}},\ }\href@noop {} {\bibfield
  {journal} {\bibinfo  {journal} {PNAS}\ }\textbf {\bibinfo {volume} {115}},\
  \bibinfo {pages} {6656} (\bibinfo {year} {2018})}\BibitemShut {NoStop}%
\bibitem [{\citenamefont {Ozawa}\ \emph {et~al.}(2020)\citenamefont {Ozawa},
  \citenamefont {Berthier}, \citenamefont {Biroli},\ and\ \citenamefont
  {Tarjus}}]{ozawa2020role}%
  \BibitemOpen
  \bibfield  {author} {\bibinfo {author} {\bibfnamefont {M.}~\bibnamefont
  {Ozawa}}, \bibinfo {author} {\bibfnamefont {L.}~\bibnamefont {Berthier}},
  \bibinfo {author} {\bibfnamefont {G.}~\bibnamefont {Biroli}}, \ and\ \bibinfo
  {author} {\bibfnamefont {G.}~\bibnamefont {Tarjus}},\ }\href@noop {}
  {\bibfield  {journal} {\bibinfo  {journal} {Physical Review Research}\
  }\textbf {\bibinfo {volume} {2}},\ \bibinfo {pages} {023203} (\bibinfo {year}
  {2020})}\BibitemShut {NoStop}%
\bibitem [{\citenamefont {Singh}\ \emph {et~al.}(2020)\citenamefont {Singh},
  \citenamefont {Ozawa},\ and\ \citenamefont {Berthier}}]{singh2020brittle}%
  \BibitemOpen
  \bibfield  {author} {\bibinfo {author} {\bibfnamefont {M.}~\bibnamefont
  {Singh}}, \bibinfo {author} {\bibfnamefont {M.}~\bibnamefont {Ozawa}}, \ and\
  \bibinfo {author} {\bibfnamefont {L.}~\bibnamefont {Berthier}},\ }\href@noop
  {} {\bibfield  {journal} {\bibinfo  {journal} {Physical Review Materials}\
  }\textbf {\bibinfo {volume} {4}},\ \bibinfo {pages} {025603} (\bibinfo {year}
  {2020})}\BibitemShut {NoStop}%
\bibitem [{\citenamefont {Yeh}\ \emph {et~al.}(2020)\citenamefont {Yeh},
  \citenamefont {Ozawa}, \citenamefont {Miyazaki}, \citenamefont {Kawasaki},\
  and\ \citenamefont {Berthier}}]{yeh2020glass}%
  \BibitemOpen
  \bibfield  {author} {\bibinfo {author} {\bibfnamefont {W.-T.}\ \bibnamefont
  {Yeh}}, \bibinfo {author} {\bibfnamefont {M.}~\bibnamefont {Ozawa}}, \bibinfo
  {author} {\bibfnamefont {K.}~\bibnamefont {Miyazaki}}, \bibinfo {author}
  {\bibfnamefont {T.}~\bibnamefont {Kawasaki}}, \ and\ \bibinfo {author}
  {\bibfnamefont {L.}~\bibnamefont {Berthier}},\ }\href@noop {} {\bibfield
  {journal} {\bibinfo  {journal} {Physical Review Letters}\ }\textbf {\bibinfo
  {volume} {124}},\ \bibinfo {pages} {225502} (\bibinfo {year}
  {2020})}\BibitemShut {NoStop}%
\bibitem [{\citenamefont {Bhaumik}\ \emph {et~al.}(2019)\citenamefont
  {Bhaumik}, \citenamefont {Foffi},\ and\ \citenamefont
  {Sastry}}]{bhaumik2019role}%
  \BibitemOpen
  \bibfield  {author} {\bibinfo {author} {\bibfnamefont {H.}~\bibnamefont
  {Bhaumik}}, \bibinfo {author} {\bibfnamefont {G.}~\bibnamefont {Foffi}}, \
  and\ \bibinfo {author} {\bibfnamefont {S.}~\bibnamefont {Sastry}},\
  }\href@noop {} {\bibfield  {journal} {\bibinfo  {journal} {arXiv preprint
  arXiv:1911.12957}\ } (\bibinfo {year} {2019})}\BibitemShut {NoStop}%
\bibitem [{\citenamefont {Varnik}\ \emph {et~al.}(2003)\citenamefont {Varnik},
  \citenamefont {Bocquet}, \citenamefont {Barrat},\ and\ \citenamefont
  {Berthier}}]{varnik2003shear}%
  \BibitemOpen
  \bibfield  {author} {\bibinfo {author} {\bibfnamefont {F.}~\bibnamefont
  {Varnik}}, \bibinfo {author} {\bibfnamefont {L.}~\bibnamefont {Bocquet}},
  \bibinfo {author} {\bibfnamefont {J.-L.}\ \bibnamefont {Barrat}}, \ and\
  \bibinfo {author} {\bibfnamefont {L.}~\bibnamefont {Berthier}},\ }\href@noop
  {} {\bibfield  {journal} {\bibinfo  {journal} {Phys. Rev. Lett}\ }\textbf
  {\bibinfo {volume} {90}},\ \bibinfo {pages} {095702} (\bibinfo {year}
  {2003})}\BibitemShut {NoStop}%
\bibitem [{\citenamefont {Shi}\ and\ \citenamefont
  {Falk}(2005)}]{shi2005strain}%
  \BibitemOpen
  \bibfield  {author} {\bibinfo {author} {\bibfnamefont {Y.}~\bibnamefont
  {Shi}}\ and\ \bibinfo {author} {\bibfnamefont {M.~L.}\ \bibnamefont {Falk}},\
  }\href@noop {} {\bibfield  {journal} {\bibinfo  {journal} {Phys. Rev. Lett}\
  }\textbf {\bibinfo {volume} {95}},\ \bibinfo {pages} {095502} (\bibinfo
  {year} {2005})}\BibitemShut {NoStop}%
\bibitem [{\citenamefont {Ketkaew}\ \emph {et~al.}(2018)\citenamefont
  {Ketkaew}, \citenamefont {Chen}, \citenamefont {Wang}, \citenamefont {Datye},
  \citenamefont {Fan}, \citenamefont {Pereira}, \citenamefont {Schwarz},
  \citenamefont {Liu}, \citenamefont {Yamada}, \citenamefont {Dmowski} \emph
  {et~al.}}]{ketkaew2018mechanical}%
  \BibitemOpen
  \bibfield  {author} {\bibinfo {author} {\bibfnamefont {J.}~\bibnamefont
  {Ketkaew}}, \bibinfo {author} {\bibfnamefont {W.}~\bibnamefont {Chen}},
  \bibinfo {author} {\bibfnamefont {H.}~\bibnamefont {Wang}}, \bibinfo {author}
  {\bibfnamefont {A.}~\bibnamefont {Datye}}, \bibinfo {author} {\bibfnamefont
  {M.}~\bibnamefont {Fan}}, \bibinfo {author} {\bibfnamefont {G.}~\bibnamefont
  {Pereira}}, \bibinfo {author} {\bibfnamefont {U.~D.}\ \bibnamefont
  {Schwarz}}, \bibinfo {author} {\bibfnamefont {Z.}~\bibnamefont {Liu}},
  \bibinfo {author} {\bibfnamefont {R.}~\bibnamefont {Yamada}}, \bibinfo
  {author} {\bibfnamefont {W.}~\bibnamefont {Dmowski}},  \emph {et~al.},\
  }\href@noop {} {\bibfield  {journal} {\bibinfo  {journal} {Nature
  communications}\ }\textbf {\bibinfo {volume} {9}},\ \bibinfo {pages} {1}
  (\bibinfo {year} {2018})}\BibitemShut {NoStop}%
\bibitem [{\citenamefont {Kapteijns}\ \emph {et~al.}(2019)\citenamefont
  {Kapteijns}, \citenamefont {Ji}, \citenamefont {Brito}, \citenamefont
  {Wyart},\ and\ \citenamefont {Lerner}}]{kapteijns2019fast}%
  \BibitemOpen
  \bibfield  {author} {\bibinfo {author} {\bibfnamefont {G.}~\bibnamefont
  {Kapteijns}}, \bibinfo {author} {\bibfnamefont {W.}~\bibnamefont {Ji}},
  \bibinfo {author} {\bibfnamefont {C.}~\bibnamefont {Brito}}, \bibinfo
  {author} {\bibfnamefont {M.}~\bibnamefont {Wyart}}, \ and\ \bibinfo {author}
  {\bibfnamefont {E.}~\bibnamefont {Lerner}},\ }\href@noop {} {\bibfield
  {journal} {\bibinfo  {journal} {Phys. Rev. E}\ }\textbf {\bibinfo {volume}
  {99}},\ \bibinfo {pages} {012106} (\bibinfo {year} {2019})}\BibitemShut
  {NoStop}%
\bibitem [{\citenamefont {Mizuno}\ \emph {et~al.}(2013)\citenamefont {Mizuno},
  \citenamefont {Mossa},\ and\ \citenamefont {Barrat}}]{mizuno2013measuring}%
  \BibitemOpen
  \bibfield  {author} {\bibinfo {author} {\bibfnamefont {H.}~\bibnamefont
  {Mizuno}}, \bibinfo {author} {\bibfnamefont {S.}~\bibnamefont {Mossa}}, \
  and\ \bibinfo {author} {\bibfnamefont {J.-L.}\ \bibnamefont {Barrat}},\
  }\href@noop {} {\bibfield  {journal} {\bibinfo  {journal} {Physical Review
  E}\ }\textbf {\bibinfo {volume} {87}},\ \bibinfo {pages} {042306} (\bibinfo
  {year} {2013})}\BibitemShut {NoStop}%
\bibitem [{\citenamefont {Shakerpoor}\ \emph {et~al.}(2020)\citenamefont
  {Shakerpoor}, \citenamefont {Flenner},\ and\ \citenamefont
  {Szamel}}]{shakerpoor2020stability}%
  \BibitemOpen
  \bibfield  {author} {\bibinfo {author} {\bibfnamefont {A.}~\bibnamefont
  {Shakerpoor}}, \bibinfo {author} {\bibfnamefont {E.}~\bibnamefont {Flenner}},
  \ and\ \bibinfo {author} {\bibfnamefont {G.}~\bibnamefont {Szamel}},\
  }\href@noop {} {\bibfield  {journal} {\bibinfo  {journal} {Soft Matter}\
  }\textbf {\bibinfo {volume} {16}},\ \bibinfo {pages} {914} (\bibinfo {year}
  {2020})}\BibitemShut {NoStop}%
\bibitem [{\citenamefont {Lerner}(2019)}]{lerner2019mechanical}%
  \BibitemOpen
  \bibfield  {author} {\bibinfo {author} {\bibfnamefont {E.}~\bibnamefont
  {Lerner}},\ }\href@noop {} {\bibfield  {journal} {\bibinfo  {journal}
  {Journal of Non-Crystalline Solids}\ }\textbf {\bibinfo {volume} {522}},\
  \bibinfo {pages} {119570} (\bibinfo {year} {2019})}\BibitemShut {NoStop}%
\bibitem [{\citenamefont {Richard}\ \emph {et~al.}(2020)\citenamefont
  {Richard}, \citenamefont {Ozawa}, \citenamefont {Patinet}, \citenamefont
  {Stanifer}, \citenamefont {Shang}, \citenamefont {Ridout}, \citenamefont
  {Xu}, \citenamefont {Zhang}, \citenamefont {Morse}, \citenamefont {Barrat}
  \emph {et~al.}}]{richard2020predicting}%
  \BibitemOpen
  \bibfield  {author} {\bibinfo {author} {\bibfnamefont {D.}~\bibnamefont
  {Richard}}, \bibinfo {author} {\bibfnamefont {M.}~\bibnamefont {Ozawa}},
  \bibinfo {author} {\bibfnamefont {S.}~\bibnamefont {Patinet}}, \bibinfo
  {author} {\bibfnamefont {E.}~\bibnamefont {Stanifer}}, \bibinfo {author}
  {\bibfnamefont {B.}~\bibnamefont {Shang}}, \bibinfo {author} {\bibfnamefont
  {S.}~\bibnamefont {Ridout}}, \bibinfo {author} {\bibfnamefont
  {B.}~\bibnamefont {Xu}}, \bibinfo {author} {\bibfnamefont {G.}~\bibnamefont
  {Zhang}}, \bibinfo {author} {\bibfnamefont {P.}~\bibnamefont {Morse}},
  \bibinfo {author} {\bibfnamefont {J.-L.}\ \bibnamefont {Barrat}},  \emph
  {et~al.},\ }\href@noop {} {\bibfield  {journal} {\bibinfo  {journal} {arXiv
  preprint arXiv:2003.11629}\ } (\bibinfo {year} {2020})}\BibitemShut {NoStop}%
\bibitem [{\citenamefont {Coslovich}(2011)}]{coslovich2011locally}%
  \BibitemOpen
  \bibfield  {author} {\bibinfo {author} {\bibfnamefont {D.}~\bibnamefont
  {Coslovich}},\ }\href@noop {} {\bibfield  {journal} {\bibinfo  {journal}
  {Physical Review E}\ }\textbf {\bibinfo {volume} {83}},\ \bibinfo {pages}
  {051505} (\bibinfo {year} {2011})}\BibitemShut {NoStop}%
\bibitem [{\citenamefont {Coslovich}\ and\ \citenamefont
  {Jack}(2016)}]{coslovich2016structure}%
  \BibitemOpen
  \bibfield  {author} {\bibinfo {author} {\bibfnamefont {D.}~\bibnamefont
  {Coslovich}}\ and\ \bibinfo {author} {\bibfnamefont {R.~L.}\ \bibnamefont
  {Jack}},\ }\href@noop {} {\bibfield  {journal} {\bibinfo  {journal} {Journal
  of Statistical Mechanics: Theory and Experiment}\ }\textbf {\bibinfo {volume}
  {2016}},\ \bibinfo {pages} {074012} (\bibinfo {year} {2016})}\BibitemShut
  {NoStop}%
\bibitem [{\citenamefont {Andersen}\ \emph {et~al.}(1971)\citenamefont
  {Andersen}, \citenamefont {Weeks},\ and\ \citenamefont
  {Chandler}}]{andersen1971relationship}%
  \BibitemOpen
  \bibfield  {author} {\bibinfo {author} {\bibfnamefont {H.~C.}\ \bibnamefont
  {Andersen}}, \bibinfo {author} {\bibfnamefont {J.~D.}\ \bibnamefont {Weeks}},
  \ and\ \bibinfo {author} {\bibfnamefont {D.}~\bibnamefont {Chandler}},\
  }\href@noop {} {\bibfield  {journal} {\bibinfo  {journal} {Physical Review
  A}\ }\textbf {\bibinfo {volume} {4}},\ \bibinfo {pages} {1597} (\bibinfo
  {year} {1971})}\BibitemShut {NoStop}%
\bibitem [{\citenamefont {Berthier}\ and\ \citenamefont
  {Tarjus}(2009)}]{berthier2009nonperturbative}%
  \BibitemOpen
  \bibfield  {author} {\bibinfo {author} {\bibfnamefont {L.}~\bibnamefont
  {Berthier}}\ and\ \bibinfo {author} {\bibfnamefont {G.}~\bibnamefont
  {Tarjus}},\ }\href@noop {} {\bibfield  {journal} {\bibinfo  {journal}
  {Physical review letters}\ }\textbf {\bibinfo {volume} {103}},\ \bibinfo
  {pages} {170601} (\bibinfo {year} {2009})}\BibitemShut {NoStop}%
\bibitem [{\citenamefont {Koumakis}\ and\ \citenamefont
  {Petekidis}(2011)}]{koumakis2011two}%
  \BibitemOpen
  \bibfield  {author} {\bibinfo {author} {\bibfnamefont {N.}~\bibnamefont
  {Koumakis}}\ and\ \bibinfo {author} {\bibfnamefont {G.}~\bibnamefont
  {Petekidis}},\ }\href@noop {} {\bibfield  {journal} {\bibinfo  {journal}
  {Soft Matter}\ }\textbf {\bibinfo {volume} {7}},\ \bibinfo {pages} {2456}
  (\bibinfo {year} {2011})}\BibitemShut {NoStop}%
\bibitem [{\citenamefont {Swallen}\ \emph {et~al.}(2007)\citenamefont
  {Swallen}, \citenamefont {Kearns}, \citenamefont {Mapes}, \citenamefont
  {Kim}, \citenamefont {McMahon}, \citenamefont {Ediger}, \citenamefont {Wu},
  \citenamefont {Yu},\ and\ \citenamefont {Satija}}]{swallen2007organic}%
  \BibitemOpen
  \bibfield  {author} {\bibinfo {author} {\bibfnamefont {S.~F.}\ \bibnamefont
  {Swallen}}, \bibinfo {author} {\bibfnamefont {K.~L.}\ \bibnamefont {Kearns}},
  \bibinfo {author} {\bibfnamefont {M.~K.}\ \bibnamefont {Mapes}}, \bibinfo
  {author} {\bibfnamefont {Y.~S.}\ \bibnamefont {Kim}}, \bibinfo {author}
  {\bibfnamefont {R.~J.}\ \bibnamefont {McMahon}}, \bibinfo {author}
  {\bibfnamefont {M.~D.}\ \bibnamefont {Ediger}}, \bibinfo {author}
  {\bibfnamefont {T.}~\bibnamefont {Wu}}, \bibinfo {author} {\bibfnamefont
  {L.}~\bibnamefont {Yu}}, \ and\ \bibinfo {author} {\bibfnamefont
  {S.}~\bibnamefont {Satija}},\ }\href@noop {} {\bibfield  {journal} {\bibinfo
  {journal} {Science}\ }\textbf {\bibinfo {volume} {315}},\ \bibinfo {pages}
  {353} (\bibinfo {year} {2007})}\BibitemShut {NoStop}%
\bibitem [{\citenamefont {Dawson}\ \emph {et~al.}(2011)\citenamefont {Dawson},
  \citenamefont {Zhu}, \citenamefont {Kopff}, \citenamefont {McMahon},
  \citenamefont {Yu},\ and\ \citenamefont {Ediger}}]{dawson2011highly}%
  \BibitemOpen
  \bibfield  {author} {\bibinfo {author} {\bibfnamefont {K.}~\bibnamefont
  {Dawson}}, \bibinfo {author} {\bibfnamefont {L.}~\bibnamefont {Zhu}},
  \bibinfo {author} {\bibfnamefont {L.~A.}\ \bibnamefont {Kopff}}, \bibinfo
  {author} {\bibfnamefont {R.~J.}\ \bibnamefont {McMahon}}, \bibinfo {author}
  {\bibfnamefont {L.}~\bibnamefont {Yu}}, \ and\ \bibinfo {author}
  {\bibfnamefont {M.}~\bibnamefont {Ediger}},\ }\href@noop {} {\bibfield
  {journal} {\bibinfo  {journal} {The Journal of Physical Chemistry Letters}\
  }\textbf {\bibinfo {volume} {2}},\ \bibinfo {pages} {2683} (\bibinfo {year}
  {2011})}\BibitemShut {NoStop}%
\end{thebibliography}%

\end{document}